\documentclass[10pt,twocolumn,english,superscriptaddress,lengthcheck,prl]{revtex4-2}
\usepackage[latin9]{inputenc}
\usepackage{color}
\usepackage{babel}
\usepackage{array}
\usepackage{verbatim}
\usepackage{amsmath}
\usepackage{amssymb}
\usepackage{graphicx}
\usepackage{esint}
\usepackage[unicode=true,
 bookmarks=false,
 breaklinks=false,pdfborder={0 0 1},backref=section,colorlinks=true]
 {hyperref}
\hypersetup{
 linkcolor=blue,filecolor=magenta,urlcolor=green}

\makeatletter

\DeclareTextSymbolDefault{\textquotedbl}{T1}

\usepackage{babel}
\renewcommand{\citet}{\cite}

\usepackage[caption=false]{subfig}

\@ifundefined{showcaptionsetup}{}{%
 \PassOptionsToPackage{caption=false}{subfig}}
\usepackage{subfig}
\makeatother

\begin{document}
\title{Superconductor vortex spectrum including Fermi arc states in time-reversal symmetric Weyl semimetals}
\author{Rauf Giwa}
\affiliation{University of Houston, Houston 77204, USA}
\author{Pavan Hosur}
\affiliation{University of Houston, Houston 77204, USA}
\affiliation{Texas Center for Superconductivity at the University of Houston, Houston
77204, USA}
\begin{abstract}
Using semiclassics to surmount the hurdle of bulk-surface inseparability, we derive the superconductor vortex spectrum in non-magnetic Weyl semimetals and show that it stems from the Berry phase of orbits made of Fermi arcs on opposite surfaces and bulk chiral modes. Tilting the vortex transmutes it between bosonic, fermionic and supersymmetric, produces periodic peaks in the density of states that signify novel nonlocal Majorana modes, and yields a thickness-independent spectrum at ``magic angles''. We propose (Nb,Ta)P as candidate materials and tunneling spectroscopy as the ideal experiment.\end{abstract}
\maketitle
Superconductor vortices are fundamentally quantum mechanical entities
with discrete energy levels whose structure encodes properties of
the parent superconductor and the normal metal. For instance, an ordinary
Fermi gas and conventional superconductivity lead to a gapped vortex
spectrum \citet{Caroli1964} while vortices in two dimensional (2D)
spinless $p+ip$ superconductors \citet{Read2000} and $s$-wave superconductors
that descend from a 2D Dirac fermion \citet{FuKaneProximity} host
zero energy states known as Majorana modes (MMs). MMs are
exotic states that equate a particle with its anti-particle. They harbor diverse
potential applications ranging from topological quantum computing \citet{KitaevQC03,Alicea2012,Beenakker2013,Elliott2015,Kouwenhoven2019,Lutchyn:2018aa,Ma2017}
and topological order \citet{Kitaev2006} to supersymmetry (SUSY)
\citet{Grover2012,Grover2014SUSY,Rahmani2015,Hsieh2016,Huang2017},
quantum chaos and holographic blackholes \citet{MaldacenaStanfordSYK,kitaev2}.
In condensed matter, they invariably appear as topologically protected zero energy bound states in topological defects such as superconductor vortices and domain walls\citet{Alicea2012,Beenakker2013,Elliott2015,FuKaneProximity,Hosur2011,Kitaev2000,Leijnse2012,Liu2018,Liu2017,Lutchyn2011,Ma2017,Mohanta2014,Mourik2012,Nadj-Perge2014,QiZhangRMP,Read2000,Rokhinson2012,Sato2010,Sato2017}. In recent years, the discovery of MMs in Fe-based superconductors with tunable band topology \citet{Zhang2018,Kong:2019we,Zhu2020,Wang333,Machida:2019wg,Liu2018,Liu:2020tn,Kong:2021wr,Wang2015a,Peng2019,Xu2016FST,Zhang:2019FeSC,Qin2019a,FTSZeemanVortices}
and the observation of superconductivity in several topological semimetals
\citet{Aggarwal:2015aa,Aggarwal:2017aa,Alidoust2017,Bachmann2017,Cai2019,Chen2016,Deng:2022wu,Gayen2016,Guguchia2017,He:2016aa,Huang:2018aa,Huang:2019aa,Kang:2015aa,LeeInturu2021,Li2018,Li2018b,Li:2017ab,Pan:2015aa,Qi:2016aa,Shvetsov2019,WANG2017425,Wang:2015aa,Xing204,vanDelft277}
have motivated an urgent quest to theoretically determine the vortex
spectrum given an arbitrary normal metal.

\begin{figure*}
\includegraphics[height=0.18\paperheight]{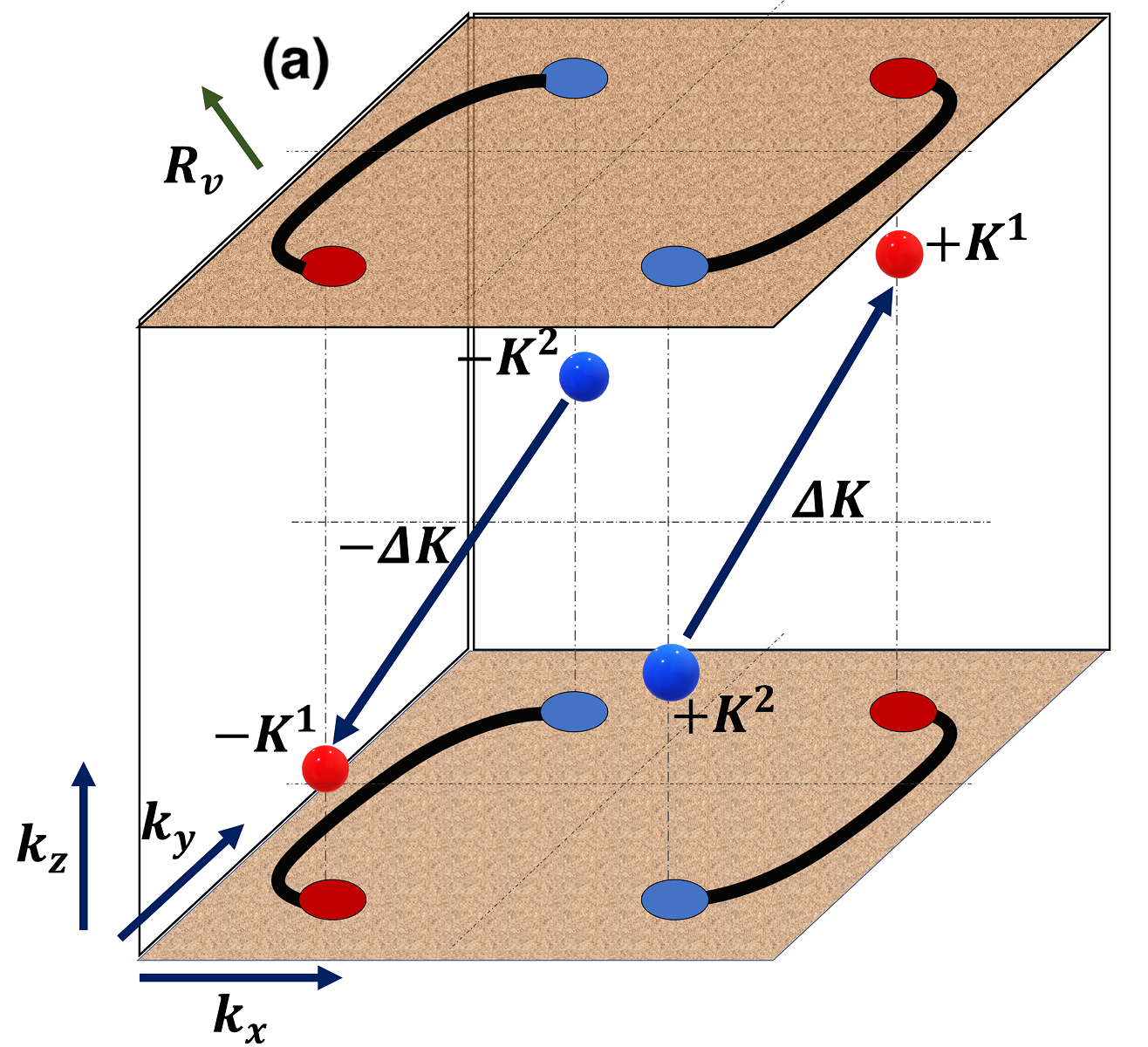}\includegraphics[height=0.18\paperheight]{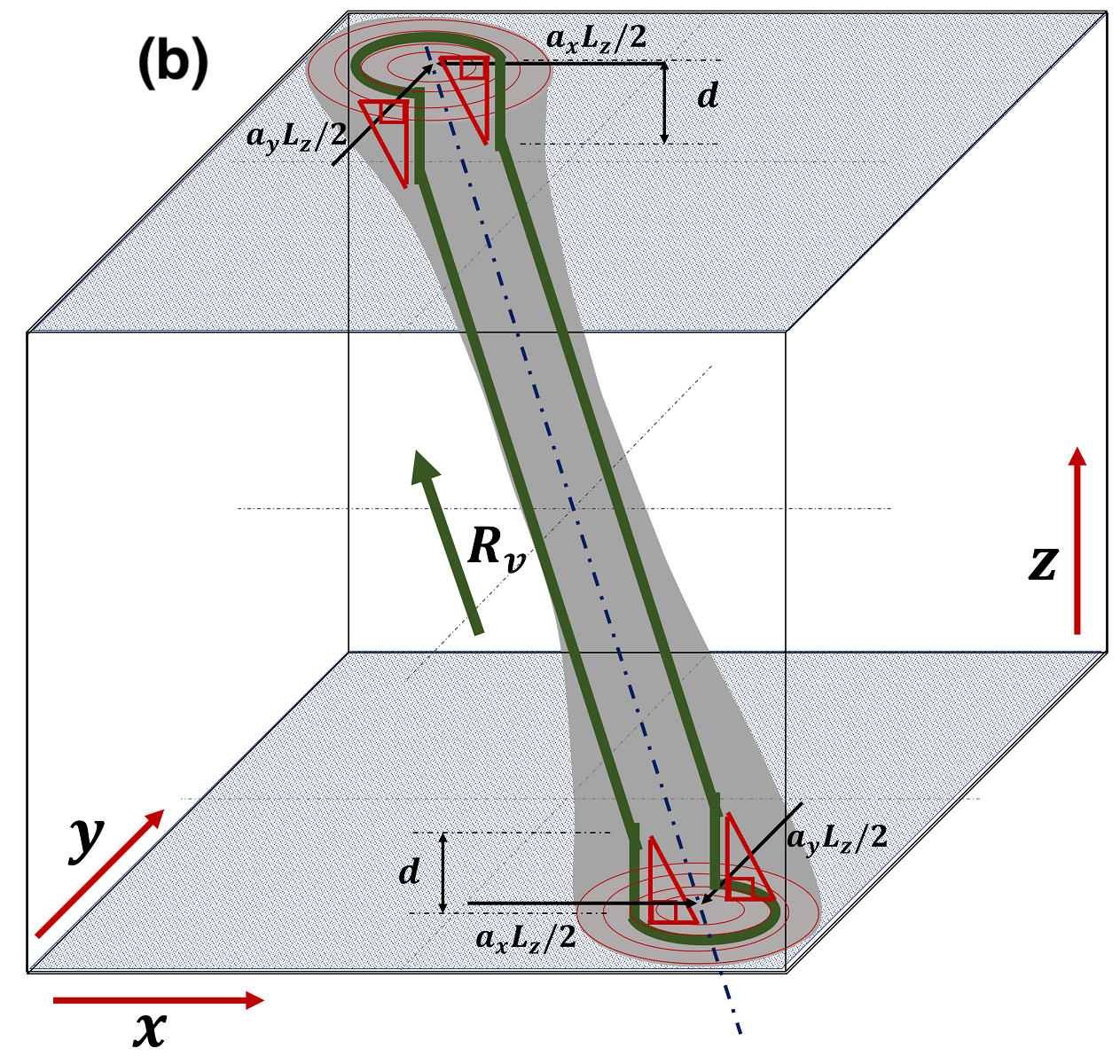}\includegraphics[height=0.18\paperheight]{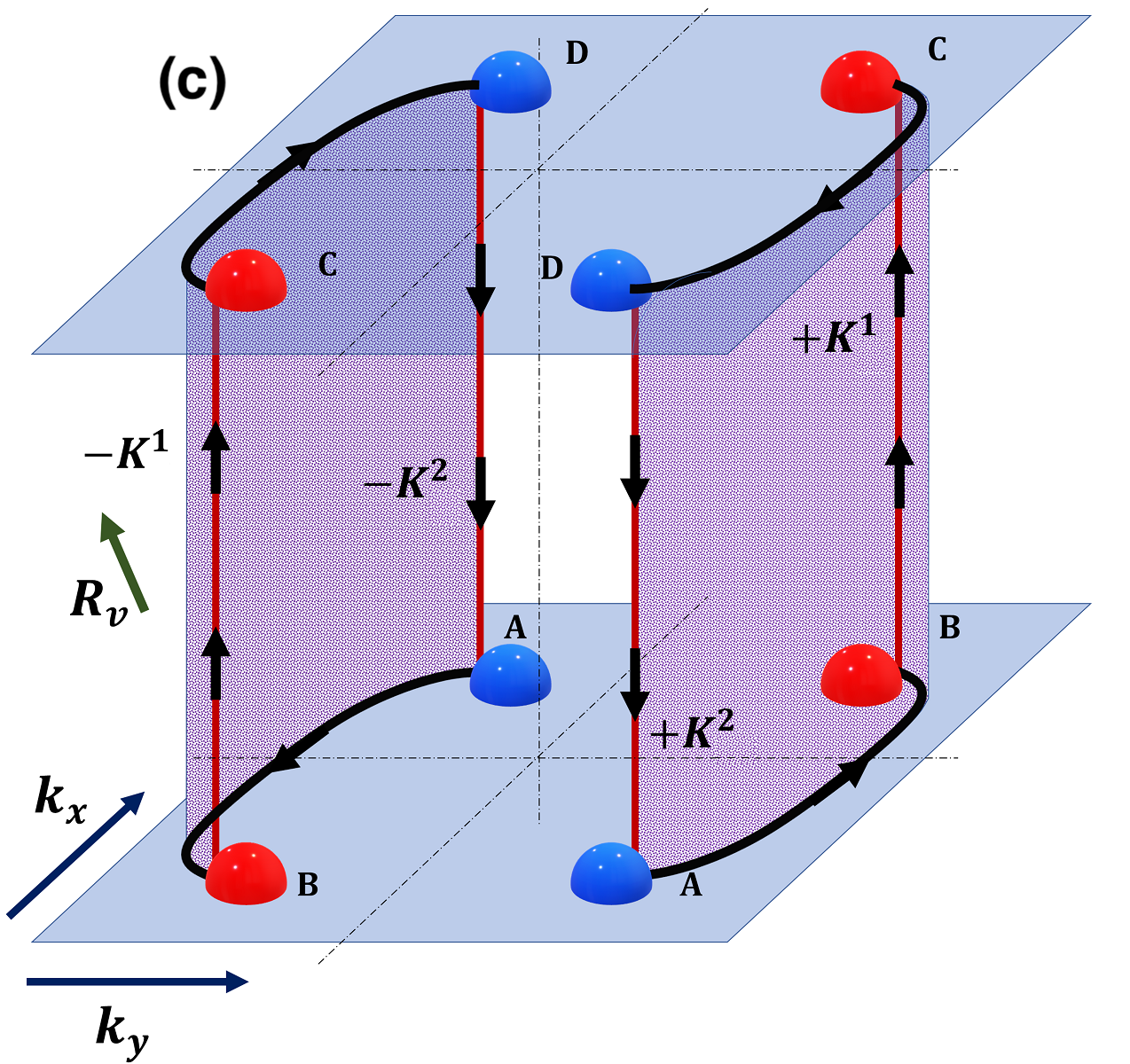}

\caption{Schematic picture. (a) $\boldsymbol{k}$-space illustration of a minimal TWSM. Red (blue) spheres at $\pm\boldsymbol{K}^{1}$
($\pm\boldsymbol{K}^{2}$) denote right-(left-) handed WNs, red (blue)
discs denote their projections onto the surface Brillouin zone, and
black curves are FAs. (b) Real space illustration of the vortex (grey
tube) and the semiclassical orbit (green curve). The classical bulk
path parallels the tube axis, but quantum tunneling causes deviations
near the surface. (c) Semiclassical orbits in mixed real ($z$) and
momentum ($k_{x},k_{y})$ space. Each orbit is a closed loop consisting of 
bulk chiral modes tied to a pair of WNs interspersed
by FAs that connect their surface projections.{\label{fig:Schematic}}}
\end{figure*}

This pursuit hits a roadblock with gapless topological
matter such as Weyl semimetals (WSMs) \citet{VafekDiracReview,Burkov2018,Burkov:2016aa,YanFelserReview,ArmitageWeylDiracReview,Shen2017,Belopolski:2016wu,Guo2018,Chang2016,Gyenis_2016,Huang:2015vn,Inoue1184,Lv:2015aa,Sun2015a,Xu2015,Xu2016,Xu613,Yang:2015aa,Zheng2016}.
In the bulk, WSMs host accidental band crossings
or Weyl nodes (WNs) that enjoy topological protection and spawn various topological responses
\citet{Hosur2013a,Wang2018,Hu:2019aa,ZyuninBurkovWeylTheta,ChenAxionResponse,VazifehEMResponse,Burkov_2015,Hosur2012,Juan:2017aa,Wang2017,Nagaosa:2020aa,NielsenABJ,IsachenkovCME,SadofyevChiralHydroNotes,Loganayagam2012,GoswamiFieldTheory,Wang2013,BasarTriangleAnomaly,LandsteinerAnomaly,Nanda2022}.
WNs carry an intrinsic chirality or handedness, and are constrained to appear in pairs of opposite chirality \citet{NielsenABJ}.
Moreover, in time-reversal ($\mathcal{T}$) symmetric WSMs (TWSMs),
each WN has a Kramer's partner of the same chirality
which leads to quadruplets of WNs. The surface of a WSM hosts Fermi
arcs (FAs) that connect
the surface projections of pairs of WNs of opposite chirality \citet{Benito-Matias2019,Chang2016,Gyenis_2016,Deng2017,Deng:2016aa,Guo2018,HaldaneFermiArc,Hosur2012a,Huang2016,Huang:2015vn,Iaia:2018aa,Inoue1184,Kwon:2020aa,Lau2017,Sakano2017,Sun2015a,Lv:2015aa,Xu2015,Xu2015a,Xu2015b,Xu2016,Xu613,XuLiu2018,Yuan2018QPI,Yuaneaaw9485,Zhang:2017ac,Moll:2016aa,Potter2014,Zhang2016},
resembling a broken segment of a 2D Fermi surface but forming a closed
loop with a FA on the opposite surface of a finite slab. The penetration
depth of a FA into the bulk depends strongly on the surface momentum
and diverges at the WN projections, thus making the surface inseparable
from the bulk. Consequently, the Fermi ``surface'' of WSMs consists
of FAs on the surface of the material and bulk Fermi points at the
WNs (or Fermi pockets around WNs not at the Fermi level). Such a Fermiology
is beyond a purely surface or purely bulk theory; yet,
a basic physical question remains: ``what is the spectrum
of a superconductor vortex in a WSM?'' 

\emph{General vortex spectrum: }We answer this question using a powerful
semiclassical approach that surmounts that above limitation. We restrict to TWSMs, since they generically host a weak pairing instability towards a gapped superconductor; WSMs that lack $\mathcal{T}$ either lack a pairing instability or yield unconventional nodal or finite-momentum pairing \cite{LiHaldane2018, ChoWeylSC, WeiWeylSC, Hao2017}. For arbitrary pairing symmetry that yields a full
gap when uniform, we propose the spectrum:

\begin{equation}
E_{n}^{\pm}=\pm\varepsilon\left(n+\dfrac{1}{2}+\dfrac{\Phi_{\mathrm{B}}+\Phi_{\mathrm{S}}-\Phi_{\text{Q}}}{2\pi}\right);\,\varepsilon=\dfrac{\Delta_{0}}{\xi l_{\mathrm{FA}}}\label{eq:vortexSpectrum}
\end{equation}
where $l_{\mathrm{FA}}$ is of order the total length of FAs on opposite surfaces that form a closed loop, $\Delta_{0}$
is the pairing amplitude far from the vortex, $\xi$ is the superconductor
coherence length and $n\in\mathbb{Z}$. Additionally, $\Phi_{\text{B}}$
is the net phase acquired by a wavepacket traversing the bulk. In
the simplest case where FAs on opposite surfaces connect the same
pairs of WNs as depicted in Fig.~\ref{fig:Schematic}, $\Phi_{\text{B}}=\Delta\boldsymbol{K}\cdot\boldsymbol{R}_{v}$
with $\Delta\boldsymbol{K}$ connecting these nodes in momentum space
and $\boldsymbol{R}_{v}$ connecting opposite ends of the vortex in
real space. Henceforth, we parameterize $\boldsymbol{R}_{v}=(a_{x}\hat{\mathbf{x}}+a_{y}\hat{\mathbf{y}}+\hat{\mathbf{z}})L_{z}\equiv(\boldsymbol{a}_{\perp}+\hat{\mathbf{z}})L_{z}$,
where $L_{z}$ is the slab thickness and $\hat{\mathbf{z}}$ is the
surface normal. Next, $\Phi_{\mathrm{S}}$ is total Berry phase of
a ``classical'' path defined by the FAs on both surfaces that ignores
their bulk penetration. Finally, the penetration effectively reduces
the thickness to $L_{z}-2d$, where $d$ is the average penetration
depth of the FAs in a region of size $O(\xi^{-1})$ around the surface projections of the Weyl nodes. This induces a quantum correction 
\begin{equation}
\Phi_{\text{Q}}=2d\Delta\boldsymbol{K}_{\perp}\cdot\boldsymbol{a}_{\perp}\label{eq:Phi-Q}
\end{equation}
where $\Delta\boldsymbol{K}_{\perp}=\Delta K_{x}\hat{\mathbf{x}}+\Delta K_{y}\hat{\mathbf{y}}$.
Thus, Eq.~(\ref{eq:vortexSpectrum}) predicts a generically non-degenerate,
discrete spectrum with equally spaced energy levels, while the zero-point
energy is determined by Berry phase of the FAs, WN locations, sample
thickness and vortex orientation. The spectrum is generically gapped,
contrary to a naive bulk approach that predicts a generically gapless spectrum
\citet{Giwa2021}.

Eq.~(\ref{eq:vortexSpectrum}) is inspired by results in Refs.~\citet{Hosur2011},
\citet{VishwanathThesis} and \citet{Giwa2021}. Ref.~\citet{VishwanathThesis}
showed that quasiparticle dynamics in inhomogeneous superconductors
can be faithfully captured by quantizing the semiclassical action
for wavepackets traveling in closed orbits in real space. The action,
which appears as a phase in the relevant path integral, was shown
to consist of three terms: (i) a Bohr-Sommerfeld phase $\oint\mathbf{k}_{\text{cl}}\cdot d\mathbf{r}_{\text{cl}}$
for the classical orbit, (ii) a Berry phase due to rotation of the
Nambu spinor, and (iii) a $\pi$ phase if a unit vortex is encircled.
Within a complementary momentum space picture, Ref.~\citet{Hosur2011}
proved that a smooth 2D Fermi surface in the normal
state and arbitrary pairing symmetry that produces a full gap
when superconductivity is uniform yield a superconductor vortex spectrum $\epsilon_{n}^{\pm}=\pm\frac{\Delta_{0}}{\xi l_{\mathrm{FS}}}\left(n+\frac{1}{2}+\frac{\Phi_{\mathrm{FS}}}{2\pi}\right)$
for $l_{\mathrm{FS}}\xi\gg1$, where $l_{\mathrm{FS}}$ and $\Phi_{\mathrm{FS}}$
are the Fermi surface perimeter and Berry phase, respectively. The
normal state is assumed to be $\mathcal{T}$-symmetric, which leads
to a pair of Fermi surfaces with opposite Berry phases in the normal
state that produce particle-hole conjugate eigenstates inside
the vortex.

To propose Eq.~(\ref{eq:vortexSpectrum}) for a TWSM, we first note
that the Bohr-Sommerfeld phase, $\pi$ phase from the vortex and the
Nambu-Berry phase contribute shifts proportional to $n$, $1/2$ and
$\Phi_{\mathrm{FS}}/2\pi$, respectively in $\epsilon_{n}$. Then,
we recall that a WN with chirality $\chi=\pm1$ produces a chiral
MM in the vortex core with chirality $\chi w$, where $w=\pm1$ is
the winding number of the vortex \citet{Giwa2021}. Thus, for $w=1$,
a right-(left-)handed WN produces a chiral MM inside the vortex with
upward (downward) group velocity. For a smooth vortex, defined by
$\left|\Delta \boldsymbol{K}\xi\right|\gg1$, these chiral modes allow
wavepackets to travel between FAs on opposite surfaces without scattering.
The smoothness also ensures that a wavepacket on the surface travels
along a single FA without scattering into other FAs. Thus, the semiclassical
orbit naturally involves travel along a FA on the top surface, tunneling
through the bulk via a downward chiral MM, FA traversal on the bottom
surface followed by tunneling up the bulk via an upward chiral MM.
Since a TWSM contains quadruplets of WNs and an even number of FAs
on each surface related by $\mathcal{T}$, such orbits appear in $\mathcal{T}$-related
pairs but with opposite energies in the vortex to preserve overall particle-hole symmetry. This picture inspires
the generalization of $\Phi_{\mathrm{FS}}$ to $\Phi_{\text{tot}}=\Phi_{\text{B}}+\Phi_{\text{S}}-\Phi_{\text{Q}}$,
the total phase acquired by a wavepacket traversing a closed orbit
in mixed real-and-momentum space, as depicted in Fig.~\ref{fig:Schematic}.

A peculiar situation occurs when $\Phi_{\text{tot}}/2\pi$ equals a half-integer. Then, Eq.~(\ref{eq:vortexSpectrum}) predicts
a gapless vortex with a pair of zero modes that can always be decomposed
into a pair of MMs in a suitable basis; see App. \ref{sec:MM-and-SUSY} for details. These MMs are highly non-local
as they are composed of mixed real-and-momentum space orbits. They
are not protected by symmetry; rather, they appear at a series of
critical points as $\Phi_{\text{tot}}$ is varied. These critical points separate trivial and topological phases of the vortex, which behaves as a 0D superconductor with a $\mathbb{Z}_2$ topological classification \cite{Chiu2016}. The MMs decouple at
criticality by definition and, when probed via an STM whose tip metal
has doubly degenerate bands, contribute separately to the tunneling
conductance. Thus, the peak height in the $dI/dV$ spectrum must be
twice that of topological MMs \citet{Law2009,Pan2021}, $2\times2e^{2}/h=4e^{2}/h$,
while the regions between critical tilts must contain quantized plateaus
separated by $4e^{2}/h$ in the $I$-$V$ characteristics.

Now, pairs of MMs separate gapped superconductors differing in fermion
parity \citet{Ran2011}. Thus, the vortex is fermionic with odd fermion parity on the topological side of
criticality, and bosonic on the trivial. Naturally, the critical
vortex is impartial to bosonic or fermionic statistics
and therefore exhibits SUSY -- a mysterious and elusive symmetry between
bosons and fermions first proposed in the Standard Model and more recently, in certain condensed matter systems \citet{Friedan1985, Grover2012,Grover2014SUSY, Qi2009, Rahmani2015,Hsieh2016,Huang2017} (see App. \ref{sec:MM-and-SUSY}
for details). Remarkably, vortices here can be tuned between
bosonic, fermionic and supersymmetric by varying $\Phi_{\text{tot}}$
which, we show below, can be accomplished by simply tilting the magnetic field that threads the vortex. While disorder, the Zeeman effect and other perturbations can modify the critical tilt angles, SUSY will persist at criticality as it is purely a property of the critical vortex and oblivious to how criticality was achieved.

In general, the vortex also contains purely bulk states that do
not involve the FAs. Firstly, WNs at the Fermi level will produce modes
$E_{\text{bulk}}^{\pm}(n_1,n_2,q_3)=\pm\sqrt{2\hbar(v_1n_1+v_2n_2)\Delta_{0}/\xi+(v_3\hbar q_3)^{2}}$,
where $n_{1,2}\in\mathbb{Z}\geq0$, $n_1+n_2\geq1$, $q_3$
is the momentum along the vortex axis measured relative to the WN
and $(v_1,v_2,v_3)$ are the canonical Weyl speeds. These modes are non-chiral
and lie above the bulk gap $E_{g}=\sqrt{2\hbar v\Delta_{0}/\xi}$,
where $v=\min(v_{1,2})$. Clearly, $E_{g}\gg\varepsilon$ if $l_\text{FA}\xi\gg\sqrt{\Delta_{0}\xi/\hbar v}\sim1$,
assuming the standard Ginzburg-Landau relation $\Delta_{0}\sim\hbar v/\xi$.
Since $l_\text{FA}\sim|\Delta\boldsymbol{K}_\perp|\leq|\Delta\boldsymbol{K}|$, the smooth vortex limit of $|\Delta \boldsymbol{K}\xi|\gg1$ is consistent with non-chiral bulk modes from undoped WNs being at parametrically higher energies.

Secondly, the bulk can also contain Fermi pockets. In the weak-pairing,
smooth vortex limit, these pockets give rise to the spectrum $E_{\text{bulk}}^{\pm}(n,q_3)=\pm\frac{\Delta_{0}}{\xi l_{\mathrm{FS}}(q_3)}\left(n+\frac{1}{2}+\frac{\Phi_{\mathrm{FS}}(q_3)}{2\pi}\right)$
with $n\in\mathbb{Z}\geq0$. Trivial Fermi pockets that do not enclose
band crossings have $\Phi_{\mathrm{FS}}(q_3)\neq\pm\pi$ $\forall q_3$
and contribute only non-chiral modes. In contrast, Fermi surfaces
enclosing WNs have $\text{\ensuremath{\Phi}}_{\text{FS}}=-\pi$ at $q_3=0$ (relative to the WN) and contribute a single $n=0$
chiral MM that combines with the FAs to form the states described
in Eq.~(\ref{eq:vortexSpectrum}), while the $n\neq0$ modes are
non-chiral. For both types of Fermi pockets, the energy scale of the
non-chiral modes $\frac{\Delta_{0}}{\xi l_{\mathrm{FS}}(q_{z})}\lesssim\varepsilon$ if $l_{\text{FS}}\gtrsim l_{\text{FA}}$.
However, these modes can be easily distinguished from those
defined in Eq.~(\ref{eq:vortexSpectrum}) by tilting the vortex,
as we discuss shortly.

Finally, the normal state bulk can contain other point or line band
crossings too which can invalidate various aspects of our results. For instance, vortices in Dirac semimetals contain a pair of counterpropagating modes for each Dirac node \cite{Qin2019, Yan2019, Konig2019}, which can hybridize and ruin the semiclassical picture. We ignore crossings beyond unit WNs because they rely on crystalline symmetries while our focus is on generic band structures with only $\mathcal{T}$ symmetry 
\citet{Zhang2019b,Vergniory2019,Tang:2019aa}.

\begin{figure*}
{\includegraphics[height=0.16\paperheight]{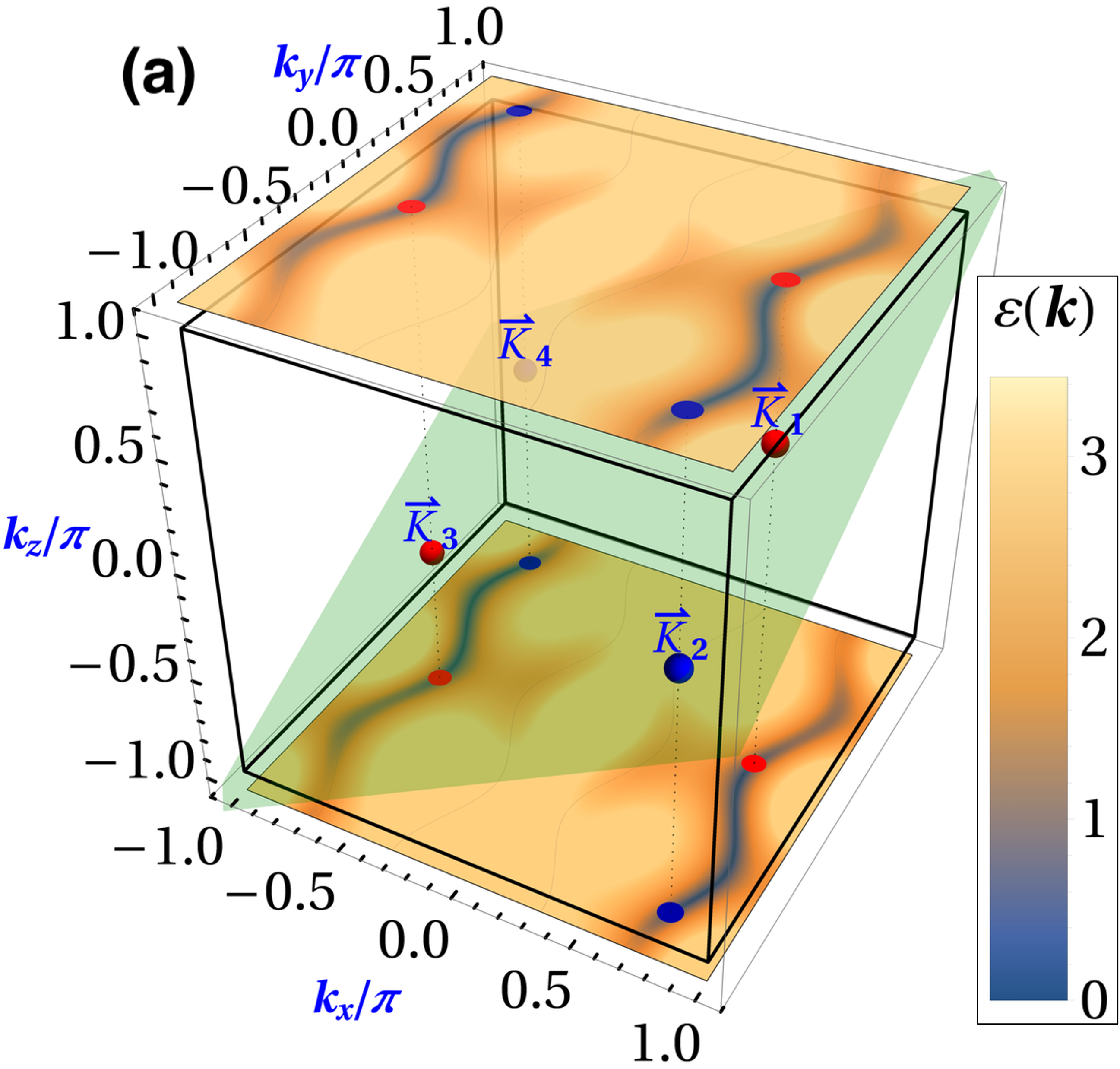}
} {\includegraphics[height=0.16\paperheight]{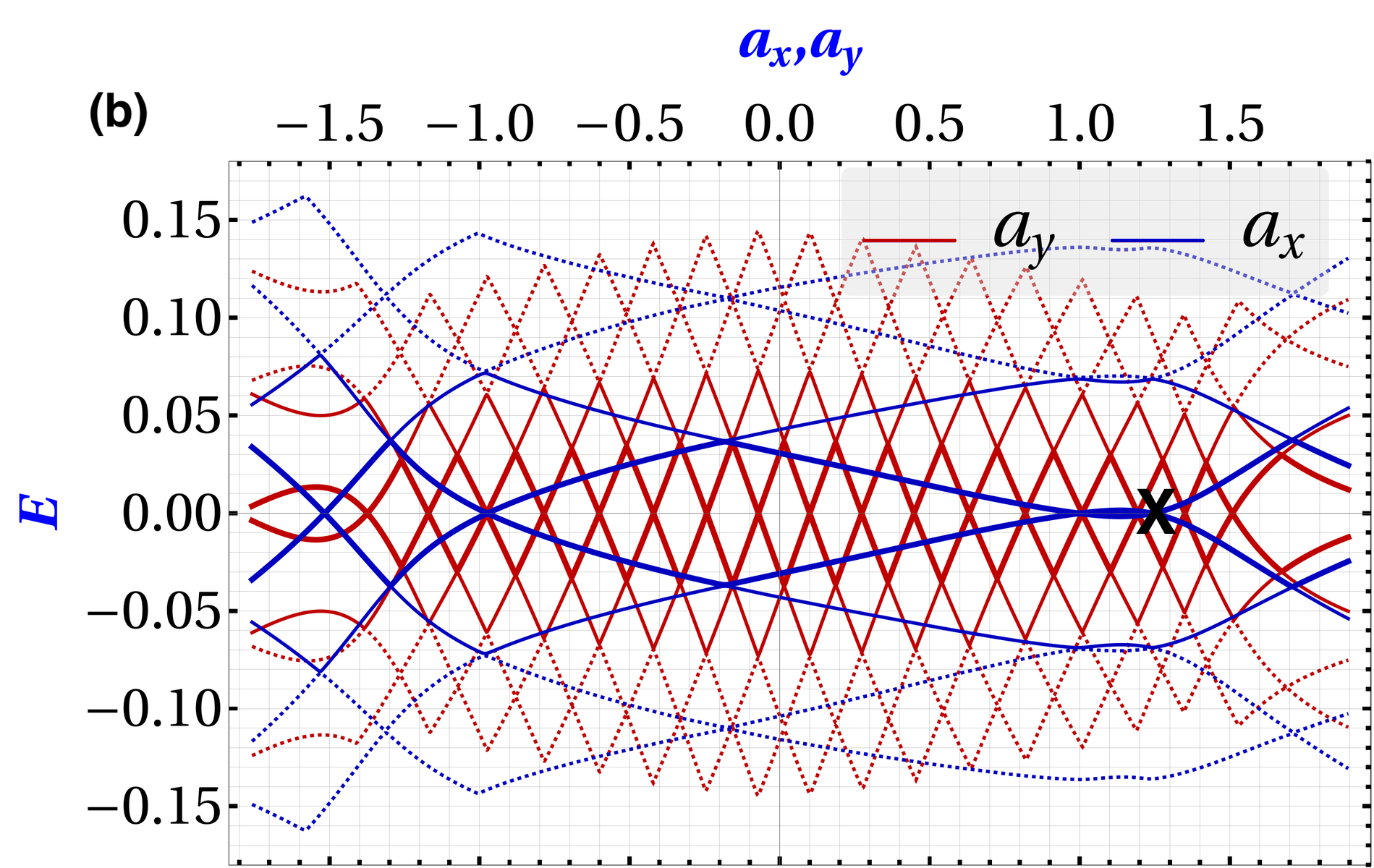}
} {\includegraphics[height=0.15\paperheight]{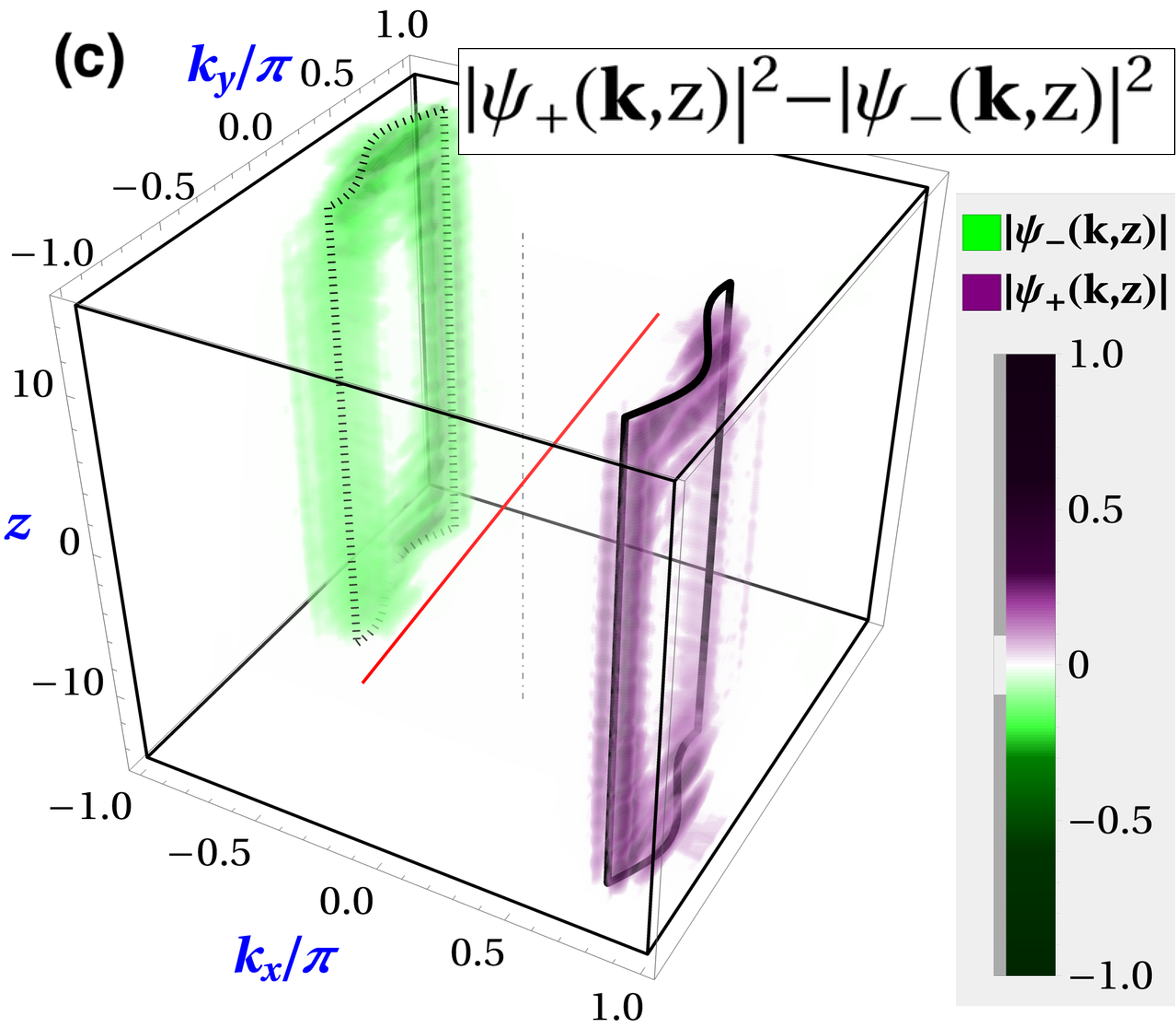}

}

\caption{Vortex spectrum for a tight-binding lattice model with unit interatomic spacing and $O(1)$ hoppings (see \citet{SM} for details). (a) Normal state band structure
showing four bulk WNs (red and blue spheres), all at different $k_{z}$,
and surface FAs connecting them. The four nodes lie on the green plane,
which is clearly not parallel to the surface. (b) Vortex spectrum
of a $L_{x}\times L_{y}\times L_{z}=23\times23\times34$ system as
the vortex is tilted separately towards the $x$-axis and the $y$-axis by $\tan^{-1}a_{i}$ ($i=x,y$).
(c) Net probability density of the two lowest energy wavefunctions
in $(k_{x},k_{y},z)$ space at $a_{y}=1.25$, marked `X' in (b), obtained
by Fourier transforming the 3D real space wavefunctions with respect
to $x,y$. We choose the band parameter $u=1.2$ which yields
$\Delta\boldsymbol{K}^{calc}=\{0.029,0.428,0.181\}\times2\pi$, and
superconducting parameters $\Delta_{0}=0.50$, $\xi=2.0$, which yield
$\varepsilon=\Delta_{0}/\xi l_{\textrm{FA}}\approx0.04$ comparable to the scale of level spacings in (b). {\label{fig:Main-results-numerics}}}
\end{figure*}

\emph{Numerical vortex spectrum:} We now support our general claims
of Eq.~(\ref{eq:vortexSpectrum}) with numerics on an orthorhombic
lattice model of a TWSM detailed in App. \ref{sec:Lattice-verification}. Given the Bloch Hamiltonian
$H_{0}(\boldsymbol{k},k_{z})$ in the normal state, the corresponding
Bogoliubov-deGennes Hamiltonian for a unit vortex along $\left(a_{x},a_{y},1\right)$
can be written as 
\begin{equation}
H_{v}=\begin{pmatrix}H_{0}\left(\boldsymbol{k},k_{z}\right) & \Delta\left(\delta\boldsymbol{r}_{\perp}\right)e^{-i\Theta\left(\delta\boldsymbol{r}_{\perp}\right)}\\
\Delta\left(\delta\boldsymbol{r}_{\perp}\right)e^{i\Theta\left(\delta\boldsymbol{r}_{\perp}\right)} & -H_{0}\left(\boldsymbol{k},k_{z}\right)
\end{pmatrix}\label{eq:tilted ScTWSM}
\end{equation}
where $\delta\boldsymbol{r}_{\perp}=(x-a_{x}z,y-a_{y}z)$, $\Theta(\delta\boldsymbol{r}_{\perp})$
is the polar angle of $\delta\boldsymbol{r}_{\perp}$ and $\Delta\left(\delta\boldsymbol{r}_{\perp}\right)=\Delta_{0}\tanh\left(|\delta\boldsymbol{r}_{\perp}|/\xi\right)$.
Direct numerical verification of Eq.~(\ref{eq:vortexSpectrum}) involves
diagonalizing $H_{v}$ in real space. However, the lack of translation
invariance in every direction limits us to relatively small $\xi$,
which causes departure from semiclassics for modest values of $n$.
We bypass this limitation by tilting the vortex and comparing the
locations of the zero modes with the predictions of Eq.~(\ref{eq:vortexSpectrum}).
This way, we always probe the lowest few energy levels, which
conform better to the semiclassical analysis. While this method allows a careful examination of the Berry phase terms and reveals various striking phenomena, $\varepsilon$ is verifiable only upto its order of magnitude.

Fig.~\ref{fig:Main-results-numerics}(a) shows the FAs and WNs in
a minimal TWSM with four WNs located at $\pm\boldsymbol{K}^{1}$ and
$\pm\boldsymbol{K}^{2}$. We chose parameters such that all nodes
are at different $k_{z}$ and $|\Delta K_{x}|\ll|\Delta K_{y}|$ where
$\Delta\boldsymbol{K}=\boldsymbol{K}^{1}-\boldsymbol{K}^{2}$. Fig.~\ref{fig:Main-results-numerics}(b)
shows the vortex spectrum for a finite slab when a vortex, initially
along $\hat{\mathbf{z}}$, is tilted separately towards the $x$-
and the $y$-axis. Tilting towards the positive $y$-axis ($a_{x}=0,a_{y}>0$)
produces numerous level crossings, which is consistent with $\Phi_{\text{B}}=(\Delta K_{y}a_{y}+\Delta K_{z})L_{z}$
changing by many multiples of $2\pi$ as $a_{y}$ varies. In contrast,
the spectrum varies weakly when the vortex is tilted towards the $x$-axis,
which is consistent with $\Phi_{\text{B}}=\Delta K_{x}a_{x}L_{z}$
varying negligibly with $a_{x}$ since $\Delta K_{x}$ itself is small.
In Fig.~\ref{fig:Main-results-numerics}(c), we plot the wavefunctions
of a pair of levels with equal and opposite energies in $(k_{x},k_{y},z)$
space. The levels, which are related by particle-hole symmetry of
the superconductor, are clearly localized around semiclassical orbits
related by $\mathcal{T}$. This confirms the picture that motivated
Eq.~(\ref{eq:vortexSpectrum}), namely, that the vortex spectrum
follows from quantizing semiclassical orbits in mixed real-and-momentum
space, and that semiclassical orbits related by $\mathcal{T}$ turn
into pairs of particle-hole conjugate quantum eigenstates. In App. \ref{sec:Lattice-verification},
we use the zero mode locations to extract $\Delta K_{y,z}$ and $\Phi_{\text{S}}$
and find remarkable agreement with expectations.

\emph{Tilting the vortex:} Besides simplifying the numerics, tilting
the vortex leads to striking qualitative phenomena. Firstly, since
$L_{z}$ enters Eq.~(\ref{eq:vortexSpectrum}) only through $\Phi_{\text{B}}$,
the spectrum becomes $L_{z}$-independent when the vortex is tilted
to a ``magic angle'' such that $\Delta\boldsymbol{K}\perp\boldsymbol{R}_{v}$
even though the semiclassical orbit still involves travel across the
bulk. Moreover, we expect peaks in the density of states, $D(E)=\sum_{n,\lambda}\delta\left(E-E_{n}^{\lambda}\right)$,
whenever $E_{n}^{\pm}=0$. Noting that $\Phi_{\text{S}}$ does not
depend on the vortex orientation, $D(0)$ peaks whenever $\Delta\boldsymbol{K}_{\perp}\cdot\boldsymbol{a}_{\perp}(L_{z}-2d)$
equals a half-integer. Thus, the tilt parameters for two successive
peaks obey 
\begin{equation}
\Delta\boldsymbol{K}_{\perp}\cdot\left[\boldsymbol{a}_{\perp}^{(j)}-\boldsymbol{a}_{\perp}^{(j+1)}\right]=\frac{2\pi}{L_{z}-2d}\label{eq:peaks}
\end{equation}
Thus, the peaks are periodic in $\boldsymbol{a}_{\perp}$ with
a period $\Delta a$ governed by the WN locations through $\Delta\boldsymbol{K}_{\perp}$
and the effective thickness, $L_{z}-2d$. Specifically, $\Delta a=\frac{2\pi}{(L_{z}-2d)\Delta K_{t}}$, where
$\Delta K_{t}$ is the component of $\Delta\boldsymbol{K}_{\perp}$
in the tilt direction.

\begin{figure}
\includegraphics[width=1.0\columnwidth]{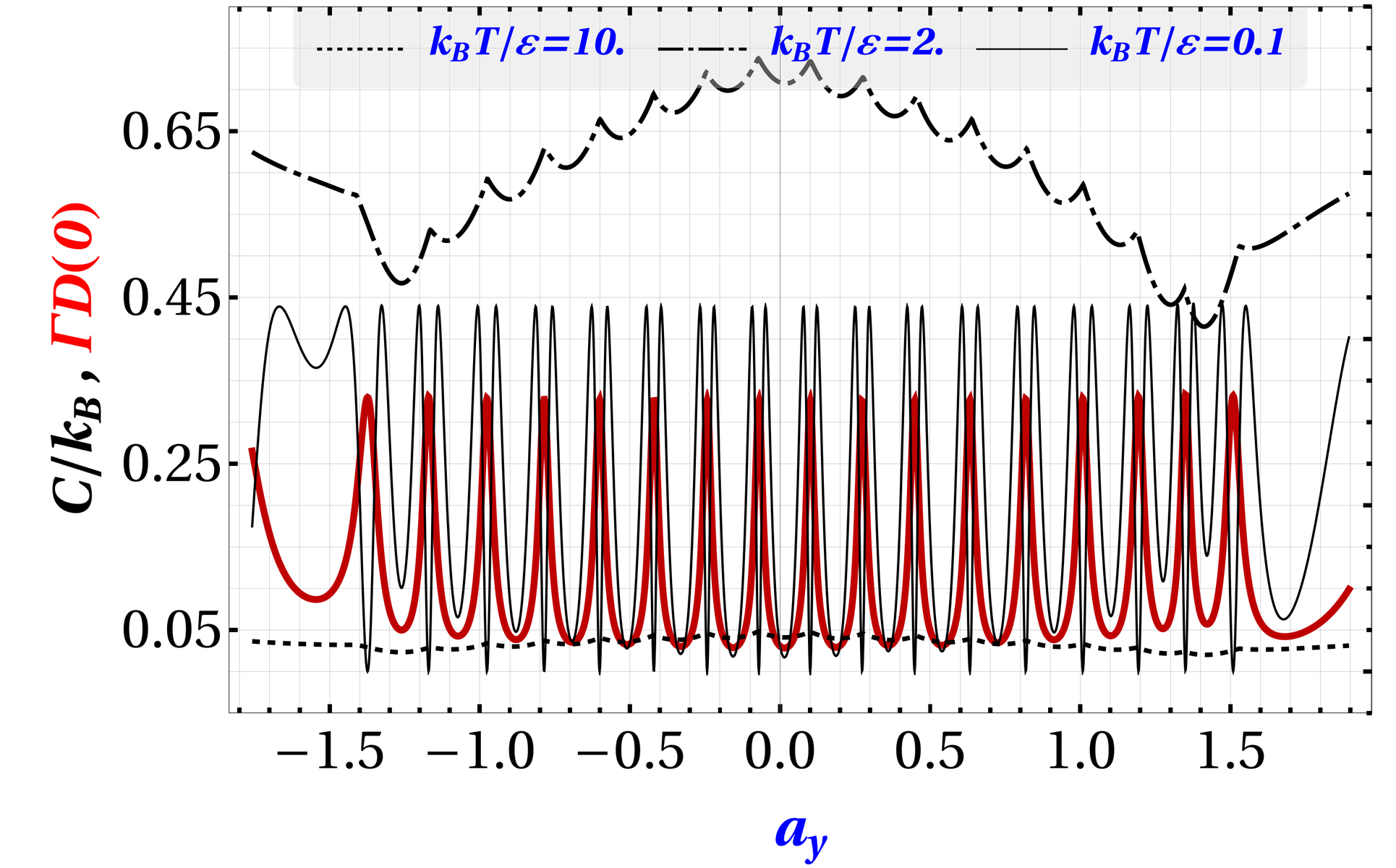}

\caption{Suitably normalized density of states $D(0)$ and specific heat $C$
at different temperatures versus $a_{y}$. Zero-modes in the spectrum
lead to sharp peaks in $D(0)$ at periodic intervals of $a_{y}$,
$\Delta a_{y}=\frac{2\pi}{(L_{z}-2d)\Delta K_{y}}$, and induce oscillations
in $C$ at low $T$ that get smeared out at high $T$. We approximate
$D(E)=\pi^{-1}\textrm{Im}\sum_{n}\left[E-E_{n}-i\Gamma\right]^{-1}$
with $\Gamma=0.0075$.\label{fig:Oscillations}}
\end{figure}

These peaks will induce characteristic oscillations with period $\Delta a$
in transport and thermodynamic quantities at temperatures below the
minigap, $T\lesssim\varepsilon/k_{B}$. For instance, the specific
heat $C=k_{B}\sum_{n}\left[\frac{E_{n}^{+}}{k_{B}T}\text{sech}\left(\frac{E_{n}^{+}}{k_{B}T}\right)\right]^{2}$
will have oscillations with a ``split-peak'' structure (Fig.~\ref{fig:Oscillations}). Similarly, a
scanning tunneling microscope (STM) should find zero bias
peaks in the differential conductance, $dI/dV$, at periodic tilts
with a peak height of $4e^{2}/h$. These oscillations can be used
to distinguish the semiclassical modes depicted in Fig.~\ref{fig:Schematic}
from non-chiral vortex modes generated by bulk Fermi pockets. The
latter are expected to produce only quantitative variations due to
the anisotropy of the Fermi pockets, but no oscillations or $L_{z}$
dependence besides finite-size effects.

The magic angle and oscillations are reminiscent of quantum oscillations
due to FAs in WSMs \citet{Borchmann2017,Potter2014,Zhang2016}. There,
a magnetic field $B$ induces cyclotron orbits involving surface FAs
and bulk chiral modes, $D(0)$ has periodic peaks in $1/B$, and $L_{z}$
enters the oscillation phase as an optical path length. Thus, at the
quantum level, the discretization predicted by Eq.~(\ref{eq:vortexSpectrum})
is analogous to Landau levels rather than finite size quantization. Indeed, if the latter was at play, Eq.~(\ref{eq:vortexSpectrum}) in the thermodynamic limit should have yielded the gapless bulk spectrum described in Ref.~\cite{Giwa2021} where FAs are irrelevant. It clearly does not, which can be attributed to the infinite penetration of the FAs into the bulk that forbids ignoring them even in this limit.
%In contrast to quantum oscillations, however, particle-hole symmetry
%of the superconductor forbids oscillations in $D(0)$ as $\Delta_{0}$
%or $\xi$ are varied. Instead, oscillations occur as $\Phi_{\text{B}}$
%and $\Phi_{\text{Q}}$ are manipulated via the vortex orientation.

\emph{Application to (Nb,Ta)P:} NbP and TaP are TWSMs in which superconductivity
induced at high pressure survives upon quenching to ambient pressure
\citet{Deng:2022wu,Li:2017ab}. Superconductivity has also been reported
in TaP directly at ambient pressure \citet{vanDelft277}. Both materials
have 24 Weyl nodes interrelated by $C_{4}$ symmetry of a face-centered
tetragonal lattice with conventional unit cell lattice constants $a_{\text{NbP}}=0.3334 \text{nm}$,
$c_{\text{NbP}}=1.1376 \text{nm}$ and $a_{\text{TaP}}=0.3318 \text{nm}$,
$c_{\text{TaP}}=1.1363 \text{nm}$ \citet{Lee2015}, and connected by
12 pairs of surface FAs. Although non-universal surface details strongly
modify the FAs and lead to non-topological gapless surface states from
trivial Fermi surfaces \citet{Sun2015a,Souma2016}, a smooth superconductor
vortex tilted in a general direction is expected to produce 12 pairs
of $\mathcal{T}$-related semiclassical orbits and hence, a superposition
of 12 different oscillations frequencies in $dI/dV$. On the other
hand, tilting in the $yz$-plane ensures that only orbits with non-zero
$\Delta K_{y}$ cause oscillations. If FAs connect surface projections
of the nearest nodes of the same family, then $\Delta K_{y}=0$ for
all 6 orbits that involve WNs separated by the $yz$-plane, while
$C_{4}^{2}$ and $\mathcal{T}$ symmetries ensure that the 6 orbits
that cross the $xz$-plane will result in precisely two frequencies:
one from WNs with $\Delta\boldsymbol{K}_{\perp,\text{NbP}}^{1}=1.0198\times\frac{2\pi}{a_{\text{NbP}}}\hat{\mathbf{y}}$
for NbP and $\Delta\boldsymbol{K}_{\perp,\text{TaP}}^{1}=0.9618\times\frac{2\pi}{a_{\text{TaP}}}\hat{\mathbf{y}}$
for TaP, and another from WNs with $\Delta\boldsymbol{K}_{\perp,\text{NbP}}^{2}=0.5406\times\frac{2\pi}{c_{\text{NbP}}}\hat{\mathbf{y}}$
and $\Delta\boldsymbol{K}_{\perp,\text{TaP}}^{2}=0.5486\times\frac{2\pi}{c_{\text{TaP}}}\hat{\mathbf{y}}$.
Discernible oscillations require $T\lesssim\varepsilon/k_{B}=\Delta_{0}/\xi l_{FA}k_{B}\sim T_{c}/\xi l_{FA}$.
Using $T_{c}\sim4K$ \citet{Deng:2022wu,Li:2017ab}, $\xi\sim4nm$
\citet{LeeInturu2021} and $l_{FA}\sim10\text{nm}^{-1}$ gives $T\lesssim0.1K$, which may be within reach
of current STM experiments. Note that $\varepsilon$ is of the same
order as the vortex minigap in typical type-II superconductors, and
STM can comfortably probe vortex modes in latter including zero bias
conductance peaks from MMs \citet{Suderow_2014,Kong:2019we,Zhu2020,Wang333,Machida:2019wg,Liu2018,Liu:2020tn,Kong:2021wr}.

In summary, we have calculated the superconductor vortex spectrum
in TWSMs including contributions from the surface FAs. While a naive
bulk calculation for a general vortex orientation suggests a gapless
spectrum consisting of a chiral mode corresponding to each WN, we
found that the low-energy spectrum is gapped in general, and determined
by the Berry phase of semiclassical orbits composed of the chiral modes and surface FAs. Such a spectrum is expected to produce a myriad
of striking phenomena upon tilting the vortex. For instance, the vortex
will alternate between bosonic and fermionic as it is tilted, while
the critical points separating the two types of vortices exhibit SUSY
and harbor unusual nonlocal MMs. Experimentally, we predict characteristic
oscillations in the specific heat and periodic, $4e^{2}/h$-quantized
peaks in the differential tunneling conductance as a function of vortex
tilt. At a certain tilt, dubbed the \textquotedbl magic angle\textquotedbl ,
the spectrum becomes independent of the slab thickness. We propose
NbP and TaP as candidate materials and tunneling spectroscopy as the
best experimental approach for studying this physics. 

\begin{acknowledgments}
We thank Liangzi Deng, Laura Greene, Kun Yang, Elio Koenig, Urjit Yajnik, Ashvin Vishwanath and Binghai Yan for valuable discussions and comments, and acknowledge financial support from the National Science
Foundation under grant DMR-2047193. 
\end{acknowledgments}

\appendix

\begin{widetext}

\section{Quantum correction}\label{sec:quantum correction}

In this section, we elaborate on the quantum correction $\Phi_{\text{Q}}$
in Eq.~1 and estimate $d$ in a tractable limit, namely, an isotropic
Weyl node with velocity $v$ at the Fermi level and a straight Fermi
arc (FA) emanating from its surface projection. These approximations
are reasonable for large $\xi$, which ensures a small momentum scale
$\sim\xi^{-1}$. Our strategy is to write down separate wavefunctions
for wavepackets deep in the bulk and on the FA, and compute their
overlap to obtain the effective distance from which a bulk wavepacket
can tunnel into the surface. We will assume the $z<0$ region to host
the Weyl fermion, measure all momenta relative to the Weyl node so
that $\boldsymbol{q}=0$ denotes the Weyl node location in the bulk
and $\boldsymbol{q}_{\perp}=(q_{x},q_{y})=\boldsymbol{0}$ denotes
its surface projection, and choose the center of the vortex on the
surface as the real space origin.

\subsection{Surface and bulk zero modes}

Suppose a FA exists in the normal state along the $q_{y}$ axis for
$q_{y}>0$. Ignoring its bulk penetration for a moment, we can model
it by the Hamiltonian
\begin{equation}
H_{\text{FA}}=\hbar vq_{x};q_{y}>0
\end{equation}
Its eigenstates are plane waves along $x$, $e^{iq_{x}x}$, for $q_{y}>0$,
while the wavefunction vanishes identically for $q_{y}<0$. Thus,
the wavefunction in this picture is discontinuous at $q_{y}=0$. 

In a time-reversal symmetric Weyl semimetal, this FA can Cooper pair
with its Kramers conjugate emanating from another Weyl node and form
a fully gapped homogenous superconductor. If the superconductor, however,
hosts an isotropic vortex, the gap amplitude must vanish at the vortex
core. Assuming a linear profile, $\Delta(\boldsymbol{r})=\Delta_{0}(x+iy)/\xi$,
the Bogoliubov-deGennes Hamiltonian in the $q_{x}>0$ region is
\begin{align}
H_{\text{FA}}^{\text{BdG}} & =\left(\begin{array}{cc}
\hbar vq_{x} & \frac{\Delta_{0}}{\xi}(x-iy)\\
\frac{\Delta_{0}}{\xi}(x+iy) & -\hbar vq_{x}
\end{array}\right)\\
 & =\Pi_{z}\hbar vq_{x}+\Pi_{x}\frac{\Delta_{0}}{\xi}x+\Pi_{y}\frac{\Delta_{0}}{\xi}y
\end{align}
where $\Pi_{x,y,z}$ are Pauli matrices in the Nambu basis. Note that
$\left[H_{\text{FA}}^{\text{BdG}},y\right]=0$, so $y=i\partial_{q_{y}}$
is a good quantum number and leads to eigenstates that are planes
waves in $q_{y}$ (analogous to plane waves in real space in systems
where momentum is a good quantum number). It is convenient to perform
a unitary rotation about $\Pi_{x}$:
\begin{align}
\tilde{H}_{\text{FA}}^{\text{BdG}} & =e^{-i\Pi_{x}\pi/4}H_{\text{FA}}^{\text{BdG}}e^{i\Pi_{x}\pi/4}\\
 & =\Pi_{z}\frac{\Delta_{0}}{\xi}y-\Pi_{y}\hbar vq_{x}+\Pi_{x}\frac{\Delta_{0}}{\xi}x\\
 & =\left(\begin{array}{cc}
\frac{\Delta_{0}}{\xi}y & i\hbar vq_{x}+\frac{\Delta_{0}}{\xi}x\\
-i\hbar vq_{x}+\frac{\Delta_{0}}{\xi}x & -\frac{\Delta_{0}}{\xi}y
\end{array}\right)
\end{align}
Integrating the energy density, $\psi^{\dagger}\tilde{H}_{\text{FA}}^{\text{BdG}}\psi$,
in $\boldsymbol{q}_{\perp}$-space across an infinetisimal region
around $q_{y}=0$ generates the boundary condition 
\begin{equation}
\psi^{\dagger}(\boldsymbol{q}_{\perp})\Pi_{z}\psi(\boldsymbol{q}_{\perp})=0\text{ at }q_{y}=0\label{eq:qx-boundary-condition}
\end{equation}

Away from this boundary for $q_{y}>0$, $\tilde{H}_{\text{FA}}^{\text{BdG}}$
is most easily solved in a Fock basis by defining bosonic creation/annihilation
operators
\begin{equation}
b=\sqrt{\frac{\xi}{2v\Delta_{0}\hbar}}\left(i\hbar vq_{x}+\frac{\Delta_{0}}{\xi}x\right)\,,\,b^{\dagger}=\sqrt{\frac{\xi}{2v\Delta_{0}\hbar}}\left(-i\hbar vq_{x}+\frac{\Delta_{0}}{\xi}x\right)
\end{equation}
which satisfy $[b,b^{\dagger}]=1$. In this basis,
\begin{equation}
\tilde{H}_{\text{FA}}^{\text{BdG}}=\left(\begin{array}{cc}
\frac{\Delta_{0}}{\xi}y & \sqrt{\frac{2\hbar v\Delta_{0}}{\xi}}b\\
\sqrt{\frac{2\hbar v\Delta_{0}}{\xi}}b^{\dagger} & -\frac{\Delta_{0}}{\xi}y
\end{array}\right)
\end{equation}
whose spectrum consists of a ``chiral mode'' on the surface that
disperses with $y$: 
\begin{align}
E_{0,y} & =\frac{\Delta_{0}}{\xi}y\label{eq:surface-chiral}\\
\psi_{0}(\boldsymbol{q}_{\perp}) & =e^{-iq_{y}y}\left(\begin{array}{c}
0\\
|0\rangle
\end{array}\right)\equiv e^{-iq_{y}y}\left(\begin{array}{c}
0\\
e^{-\frac{1}{2}q_{x}^{2}\xi_{0}^{2}}
\end{array}\right)\nonumber 
\end{align}
and non-chiral modes:%
\begin{align}
E_{m,y}^{\pm} & =\pm\sqrt{\frac{2\hbar v\Delta_{0}}{\xi}}\sqrt{m+\frac{\Delta_{0}}{2\hbar v\xi}y^{2}}\,;\,m\in\mathbb{Z}>0\\
\psi_{m}^{\pm}(\boldsymbol{q}_{\perp}) & =e^{-iq_{y}y}\left(\begin{array}{c}
\left[\sqrt{\frac{\Delta_{0}}{2\hbar v\xi}}y\pm E_{m,y}^{\pm}\right]|m\rangle\\
\sqrt{m}|m-1\rangle
\end{array}\right)\nonumber 
\end{align}
The non-chiral modes at $\pm y$ can be superposed to fulfil the boundary
condition (\ref{eq:qx-boundary-condition}). However, the surface
chiral mode (\ref{eq:surface-chiral}) is non-degenerate and cannot
satisfy the boundary condition. This is a reflection of the fact that
the surface chiral mode evolves into the bulk chiral mode as $q_{y}\to0$,
so it cannot fulfil the boundary conditions by itself. Thus, we will
manually cut it off at $q_{y}=0$ below.

So far, we have ignored the bulk penetration of the surface states.
In the presence of uniform superconductivity $\Delta$, the bulk develops
a momentum dependent gap $\sqrt{(\hbar vq)^{2}+\Delta^{2}}$. If the
surface hosts a zero mode, its penetration depth $\kappa^{-1}$ is
given by the solution to $(\hbar vq_{\perp})^{2}-(\hbar v\kappa)^{2}+\Delta^{2}=0$
which yields $\kappa=\sqrt{q_{\perp}^{2}+(\Delta/\hbar v)^{2}}$.
Clearly, $\kappa$ reduces to $q_{\perp}$ when $q_{\perp}\gg|\Delta/\hbar v|$,
i.e., for points on the FA that are sufficiently far from the Weyl
node projection. Thus, we attach such a penetration profile to the
surface chiral mode in Eq.~(\ref{eq:surface-chiral}) and write the
wavefunction of the $E_{0,y}=0$ state as
\begin{equation}
\psi_{\text{surf}}(\boldsymbol{q}_{\perp},z)=\frac{1}{\sqrt{2}}\Theta(q_{y})\exp\left[q_{\perp}z-\frac{1}{2}q_{x}^{2}\xi_{0}^{2}\right]\left(\begin{array}{c}
i\\
1
\end{array}\right)
\end{equation}
in the region $q_{\perp}\gg\Delta/\hbar v$, where $\xi_{0}=\sqrt{\hbar v\xi/\Delta_{0}}$
and we have undone the $e^{i\Pi_{x}\pi/4}$ rotation that turned $H_{\text{FA}}^{\text{BdG}}$
into $\tilde{H}_{\text{FA}}^{\text{BdG}}$. Physically, the above
form simulates the statement that states with large momentum are sensitive
to the short distance behavior of the pair potential. Since $\Delta(\boldsymbol{r})$
vanishes at the vortex core, the bulk penetration of these states
is effectively blind to the superconductivity. Put differently, wavefunctions
at different $\boldsymbol{q}_{\perp}$ are independent of one another
when the pairing is uniform; when the pairing depends on $(x,y)$,
the wavefunctions, including their bulk tails, are smoothly stitched
together by the $\boldsymbol{q}_{\perp}$-space derivatives.

Next, consider a wavepacket in the bulk that rides the chiral Majorana
mode. For a linear vortex profile, the exact eigenfunction of the
chiral Majorana mode deep in the bulk is given by a plane wave along
the vortex axis and a Gaussian of width $\xi_{0}$ in the transverse
direction \citet{Giwa2021}. We construct a wavepacket centered at
depth $|Z|\gg\xi_{0}$ whose spread along the vortex axis is also
$\xi_{0}$. Such a wavepacket is spherically symmetric, which ensures
its validity for any vortex orientation. Its wavefunction is of the
form
\begin{equation}
\psi_{\text{bulk}}(\boldsymbol{r})\propto\exp\left(-\frac{|\boldsymbol{r}-\boldsymbol{a}Z|^{2}}{2\xi_{0}^{2}}\right)
\end{equation}
where $\boldsymbol{a}=(\boldsymbol{a}_{\perp},1)$ parametrizes the
vortex orientation. Equivalently, Fourier transforming with respect
to $(x,y)$ gives
\begin{equation}
\psi_{\text{bulk}}(\boldsymbol{q}_{\perp},z)\propto\exp\left(-i\boldsymbol{q}_{\perp}\cdot\boldsymbol{a}_{\perp}Z-\frac{1}{2}q_{\perp}^{2}\xi_{0}^{2}-\frac{(z-Z)^{2}}{2\xi_{0}^{2}}\right)
\end{equation}

\subsection{Hybridization between surface and bulk modes}

The hybridization is now straightforward to calculate. The spinor
parts of $\psi_{\text{surf}}$ and $\psi_{\text{bulk}}$ will yield
an $O(1)$ matrix element in general that depends on the details of
the boundary conditions at $z=0$. The effective penetration depth
is given by the spatial part of the overlap. Explicitly, the spatial
overlap is
\begin{align}
\left\langle \psi_{\text{surf}}|\psi_{\text{bulk}}\right\rangle  & \propto\intop_{q_{\perp}>1/\xi_{0}}\intop_{z}\Theta(q_{y})\Theta(-z)\exp\left(q_{\perp}z-\xi_{0}^{2}q_{x}^{2}-\frac{1}{2}\xi_{0}^{2}q_{x}^{2}-i\boldsymbol{q}_{\perp}\cdot\boldsymbol{a}_{\perp}Z-\frac{(z-Z)^{2}}{2\xi_{0}^{2}}\right)\\
 & \approx\sqrt{2\pi}\xi_{0}\intop_{q_{\perp}>1/\xi_{0}}\Theta(q_{y})\exp\left(q_{\perp}Z-\xi_{0}^{2}q_{x}^{2}-\frac{1}{2}\xi_{0}^{2}q_{y}^{2}-i\boldsymbol{q}_{\perp}\cdot\boldsymbol{a}_{\perp}Z\right)
\end{align}
approximating $\exp\left[(Z-z)^{2}/2\xi_{0}^{2}\right]\approx\sqrt{2\pi}\xi_{0}\delta(z-Z)$.
We can further simplify the integral by dropping the quadratic terms
in the exponent under the assumption $|Z|/\xi_{0}\gg q_{\perp}\xi_{0}$.
This is reasonable as $|Z|$ has no upper bound in the thermodynamic
limit whereas $q_{\perp}$ is ultimately bounded by the inverse lattice
constant. Parametrizing $\boldsymbol{q}_{\perp}=q_{\perp}(\cos\theta,\sin\theta)$
and $\boldsymbol{a}_{\perp}=a_{\perp}(\cos\alpha,\sin\alpha)$ allows
a tractable $q_{\perp}$-integral:
\begin{align}
\left\langle \psi_{\text{surf}}|\psi_{\text{bulk}}\right\rangle  & \propto\intop_{0}^{\pi}d\theta\intop_{1/\xi_{0}}^{\infty}q_{\perp}dq_{\perp}\exp\left[q_{\perp}Z\left(1-i\cos(\theta-\alpha)\right)\right]\\
 & \approx\frac{1}{\xi_{0}^{2}}\frac{\exp\left[-|Z/\xi_{0}|\right]}{|Z/\xi_{0}|}\intop_{0}^{\pi}d\theta\frac{\exp\left[i|Z/\xi_{0}|\cos(\theta-\alpha)\right]}{1-i\cos(\theta-\alpha)}\\
 & =\frac{1}{\xi_{0}^{2}}\frac{\exp\left[-|Z/\xi_{0}|\right]}{|Z/\xi_{0}|}\intop_{-\cos\alpha}^{\cos\alpha}\frac{\text{sgn}(\zeta)d\zeta}{\sqrt{1-\zeta^{2}}}\frac{\exp\left[i|Z/\xi_{0}|\zeta\right]}{1-i\zeta}
\end{align}
where we have defined $\zeta=\cos(\theta-\alpha)$ in the last line.
The $\zeta$-integrand is a product of a highly oscillatory function
$e^{i|Z/\xi_{0}|\zeta}$ and a piecewise smooth function. Such integrals
can be approximated as follows. 

Suppose $I[f]=\intop_{a}^{b}f(x)e^{iNx}dx$ where $f(x)$ is smooth
for $x\in(a,b)$. Integrating over $d(iNx)$ by parts gives
\begin{align}
I[f] & =\frac{1}{iN}\left[f(x)e^{iNx}\right]_{a}^{b}-\frac{1}{iN}I[f']
\end{align}
For large $N$, this enables a recursive evaluation of $I[f]$:
\begin{align}
I[f] & =\frac{1}{iN}\left[f(x)e^{iNx}\right]_{a}^{b}-\frac{1}{iN}I[f']\\
 & =\frac{1}{iN}\left[f(x)e^{iNx}\right]_{a}^{b}-\frac{1}{iN}\left(\frac{1}{iN}\left[f'(x)e^{iNx}\right]_{a}^{b}-\frac{1}{iN}I[f'']\right)
\end{align}
and so on. Thus, to leading order in $1/N$,
\begin{equation}
I[f]=\frac{f(b)e^{iNb}-f(a)e^{iNa}}{iN}+O\left(\frac{1}{N^{2}}\right)
\end{equation}

Applying this result to the $\zeta$-integral, we finally get%

\begin{align}
\left\langle \psi_{\text{surf}}|\psi_{\text{bulk}}\right\rangle  & \propto\frac{\exp\left[-|Z/\xi_{0}|\right]}{|Z/\xi_{0}|^{2}}\text{sgn}(\cos\alpha)\left\{ 1-\frac{1}{|\sin\alpha|}\text{Re}\left(\frac{e^{-i|Z/\xi_{0}|\cos\alpha}}{1+i\cos\alpha}\right)\right\} 
\end{align}
This allows us to define an effective penetration depth $d$ as,
\begin{align}
\frac{1}{d} & =-\frac{1}{|Z|}\ln\left|\left\langle \psi_{\text{surf}}|\psi_{\text{bulk}}\right\rangle \right|\\
 & =\frac{1}{\xi_{0}}\left(1+\frac{2\ln|Z/\xi_{0}|}{|Z/\xi_{0}|}-\frac{\ln\left|1-\frac{1}{|\sin\alpha|}\text{Re}\left(\frac{e^{-i|Z/\xi_{0}|\cos\alpha}}{1+i\cos\alpha}\right)\right|}{|Z/\xi_{0}|}\right)
\end{align}
Recall that $\alpha$ is the polar angle, $\boldsymbol{a}_{\perp}=a_{\perp}(\cos\alpha,\sin\alpha)=(a_{x},a_{y})$,
the FA was assumed to be along $q_{x}=0;q_{y}>0$, and $\xi_{0}=\sqrt{v\xi/\Delta_{0}}\approx\xi$
within the Ginzburg-Landau theory. 

For general $\alpha$ and large $|Z|$, we see that $d$ simply equals
$\xi_{0}$, reflecting the fact that the bulk wavepacket has a spherically
symmetric spread $\sim\xi_{0}$, so it starts touching the surface
when its guiding center is within $\xi_{0}$ of the surface. In the
weak pairing, smooth vortex limit, this leads to an $O(1)$ contribution
$\Phi_{\text{Q}}$ to the total Berry phase in Eq.~1 of the main
paper. The power law factor $1/|Z/\xi_{0}|^{2}$ in the overlap $\left\langle \psi_{\text{surf}}|\psi_{\text{bulk}}\right\rangle $
gives a logarithmic correction that vanishes as $|Z/\xi_{0}|\to\infty$.
Interestingly, find a correction that diverges as $\propto\ln\left|\sin\alpha\right|$
as $\alpha\to0,\pi$ for any finite $|Z|$. At these values of $\alpha$,
the vortex axis and hence, the velocity of the bulk chiral mode are
in the $xz$ plane. Since the normal state FA disperses along $x$
for electrons or $-x$ when viewed as a dispersion of holes, this
divergence is indicative of an intuitive behavior: the bulk chiral
Majorana mode can effortlessly tunnel into the surface modes when
it does not have to change the direction of its in-plane velocity.%

In our lattice numerics, we tilt the vortex in fixed planes containing
the $z$-axis. As a result, $\alpha$ is constant for each set of
tilt-dependent data, so the divergence does not hamper the numerics.
In fact, most of our analysis involves tilting the vortex in the $yz$-plane
with the FAs are roughly parallel to $q_{y}$, so $\alpha\approx\pm\pi/2$
and $\ln|\sin\alpha|$ is negligible.

\section{Majorana modes and tunable supersymmetry\label{sec:MM-and-SUSY}}

\subsection{Nonlocal Majorana modes}

Majorana modes (MMs) in condensed matter invariably appear as topologically
protected localized zero energy bound states trapped in topological
defects such as superconductor vortices and domain walls \citet{Alicea2012,Beenakker2013,Elliott2015,FuKaneProximity,Hosur2011,Kitaev2000,Leijnse2012,Liu2018,Liu2017,Lutchyn2011,Ma2017,Mohanta2014,Mourik2012,Nadj-Perge2014,QiZhangRMP,Read2000,Rokhinson2012,Sato2010,Sato2017}.
We now show that the semiclassical orbits described here give rise
to a novel class of MMs that are nonlocal in mixed real-and-momentum
space. 

As depicted in Fig.~1(c) and computed numerically in Fig.~2(c) of
the main paper, semiclassical orbits in a TWSM appear in pairs related
by $\mathcal{T}$ and guarantee a particle-hole symmetric spectrum.
Specifically, the two orbits in a pair yield quantum eigenstates $|n\pm\rangle$
with opposite energies $E_{n}^{\pm}$ such that 
\begin{equation}
\mathcal{C}|n,\lambda\rangle=|n,-\lambda\rangle\,,\,H_{v}|n,\lambda\rangle=E_{n}^{\lambda}|n,\lambda\rangle;\lambda=\pm
\end{equation}
where $\mathcal{C}$ denotes charge conjugation and $H_{v}$ is the
vortex Bogoliubov-deGennes Hamiltonian. Whenever $E_{n}^{\lambda}=0$,
``cat'' superpositions of the eigenstates are simultaneous eigenstates
of $H_{v}$ and $\mathcal{C}$: 
\begin{equation}
\mathcal{C}\left(\frac{|n,+\rangle\pm|n,-\rangle}{\sqrt{2}}\right)=\pm\left(\frac{|n,+\rangle\pm|n,-\rangle}{\sqrt{2}}\right)
\end{equation}
As a result, $|\chi_{n,\pm}\rangle=\frac{1}{\sqrt{2}}\left(|n,+\rangle\pm|n,-\rangle\right)\equiv\chi_{n,\pm}|\phi\rangle$
are MMs, where $|\phi\rangle$ denotes particle vacuum and $\chi_{n,\pm}$
are Majorana operators that obey $\chi_{n,\lambda}^{\dagger}=\chi_{n,\lambda}$
and ${\chi_{n,\lambda},\chi_{n,\lambda'}}=\delta_{\lambda,\lambda'}$.
For general band and vortex parameters, the vortex belongs to class
D in the Altland-Zirnbauer classification as $\mathcal{C}^{2}=+1$
while $\mathcal{T}$ symmetry is broken. Since the finite thickness
is crucial to the physics described here, the vortex is effectively
a 0D superconductor characterized by a $\mathbb{Z}_{2}$ topological
invariant, $\nu\in\{0,1\}$, where the trivial and topological phases
correspond to even and odd fermion parity \citet{Chiu2016}. These
0D ``phases'' are separated by the critical MMs $|\chi_{n,\pm}\rangle$.

\subsection{Tunable SUSY and vortex statistics}

Supersymmetry (SUSY) is a symmetry between matter/fermionic and force/bosonic
particles that was originally proposed as an extension of the Standard
Model. While it has received little experimental support in particle
physics, several condensed matter systems have been shown to exhibit
SUSY including the 1D Ising model at the tricritical point \citet{Friedan1985},
boundaries \citet{Grover2012,Grover2014SUSY} and defects \citet{Qi2009}
in topological superconductors and chains and arrays of interacting
MMs \citet{Rahmani2015,Hsieh2016,Huang2017}. Unfortunately, these
proposals face serious practical challenges such as the inability
to tune the necessary parameters dynamically and the absence of materials
that realize the parent topological phases, thus rendering SUSY experimentally
elusive. We now argue that the critical points discussed earlier,
remarkably, exhibit SUSY, and the fundamental exchange statistics
of the vortices -- bosonic vs fermionic -- can be toggled by simply
tilting them across these critical points. 

Intuitively, a vortex is bosonic (fermionic) if it is effectively
a 0D superconductor with even (odd) fermion parity. In this context,
SUSY is essentially the degeneracy between bosonic and fermionic vortices
at criticality. In particular, $\chi_{n,\pm}$ form a complex fermion
state that can be either occupied or unoccupied, resulting in vortices
with distinct fermion parity at the same many-body energy. For fixed
tilt, no local measurement will be able to distinguish between bosonic
and fermionic vortices. The current realization should be more experimentally
accessible than previous proposals as it relies on existing phases
of matter, namely, TWSMs and conventional type-II superconductors,
and a simple tuning parameter, namely, magnetic field direction. In
fact, in any real material, natural variations in vortex orientations
will likely result in both bosonic and fermionic vortices, making
it the only system to the best of our knowledge where the same type
of excitation appears as both bosons and fermions.

To see the SUSY explicitly, we write the vortex Hamiltonian and many-body
ground state in second quantized form as $H_{v}=\sum_{m}E_{m}^{+}c_{m,+}^{\dagger}c_{m,+}-\mathcal{E}$
and $|G\rangle=\prod_{m}c_{m,\textrm{sgn}(E_{m}^{-})}^{\dagger}|\phi\rangle$,
respectively, where the fermion operators obey $c_{m,+}^{\dagger}=c_{m,-}$
in the physical Hilbert space and the constant $\mathcal{E}=-\sum_{m}|E_{m}^{+}|$
ensures $H_{v}$ is non-negative definite. When a single-particle
energy $E_{n}^{+}=0$, the corresponding fermion operators get promoted
to symmetries: $[H_{v},c_{n,+}]=[H_{v},c_{n,+}^{\dagger}]=0$. This
allows us to introduce operators 
\begin{equation}
Q=c_{n,+}\sqrt{H_{v}}\;,\;Q^{\dagger}=c_{n,+}^{\dagger}\sqrt{H_{v}}
\end{equation}
that obey the superalgebra 
\begin{equation}
\{Q,Q^{\dagger}\}=H_{v}\;,\;\{Q,Q\}=\{Q^{\dagger},Q^{\dagger}\}=0
\end{equation}
and hence, define an $\mathcal{N}=2$ SUSY in 0D.

\subsection{Majorana modes and SUSY at the magic angle \label{subsec:MM-SUSY-magic}}

In real materials, the detailed shape and connectivity of FAs depends
on the local boundary conditions. Nonetheless, significant insight
can be obtained by considering a slab of a material exposed to vacuum
on either side. This approach is routinely adopted in theoretical
treatments of topological phases and is pertinent when the topological
material either couples weakly to the substrates or is sandwiched
between identical substrates. In WSMs, a feature that frequently appears
in this limit is exact overlap between FAs on the top and the bottom
surfaces. We now show that magic angle vortices in materials and models
with this feature are critical, provided an additional symmetry condition,
detailed below, is satisfied. This result should facilitate the search
for parameters and materials where the vortex is critical, and therefore
carries nonlocal MMs and exhibits SUSY.

We first describe the conditions that protect the coincidence of FAs
on opposite surfaces in generic TWSMs. Suppose an operation $\mathcal{P}$
relates the coinciding FAs: 
\begin{equation}
\mathcal{P}|t(\boldsymbol{k})\rangle=e^{i\eta(\boldsymbol{k})}|b(\boldsymbol{k})\rangle;\mathcal{P}|b(\boldsymbol{k})\rangle=e^{i\eta^{\prime}(\boldsymbol{k})}|t(\boldsymbol{k})\rangle\label{eq:P-top-bottom-1}
\end{equation}
where $|t(\boldsymbol{k})\rangle$, $|b(\boldsymbol{k})\rangle$ denote
the exact FA states on the top and bottom surfaces respectively including
their spatial profile in $z$, $\boldsymbol{k}=(k_{x},k_{y})$ is
the in-plane momentum and $\eta(\boldsymbol{k}),\eta^{\prime}(\boldsymbol{k})$
are phases. $\mathcal{P}$ must change $z\to-z$ to interchange the
FAs, must preserve $\boldsymbol{k}$ so that it can protect overlap
between FAs of arbitrary shape, and must not change $k_{z}$ to preserve
the locations of the bulk Weyl nodes. This restricts $\mathcal{P}$
to be anti-unitary and of the form $\mathcal{P}=\mathcal{T}\tilde{I}$,
where $\tilde{I}$ denotes spatial inversion followed by a local unitary
transformation within a unit cell. Next, suppose $[\mathcal{P},\hat{H}(\boldsymbol{k})]=0$,
where $\hat{H}(\boldsymbol{k})$ is the Bloch Hamiltonian matrix at
$\boldsymbol{k}$. $\mathcal{P}$ would then conserve energy and cause
the FAs it relates to disperse in the same direction. However, this
leads to a contradiction as each point on the FA contour constitutes
an edge state of a 2D Chern insulator defined on a momentum-space
sheet that encloses a Weyl node and edge states of Chern insulators
must necessarily disperse in opposite directions. If $\{\mathcal{P},\hat{H}(\boldsymbol{k})\}=0$
instead, $\mathcal{P}$ can protect the coincidence of FAs at zero
energy and allow them to disperse in opposite directions. The upshot
is that protected overlap between FAs on opposite surfaces in generic
TWSMs requires a particle-hole symmetry at each $\boldsymbol{k}$.
We now show that if $\mathcal{P}^{2}=-1$, so that $\eta^{\prime}(\boldsymbol{k})-\eta(\boldsymbol{k})=\pi$
and $\hat{H}\in$ CII in the Altland-Zirnbauer classification, a striking
phenomenon occurs: zero modes exist in the vortex spectrum precisely
at the magic angle, as seen in Fig.~\ref{fig:fits}(a) at $a_{y}\approx0.424$.

At the magic angle, $\Delta\boldsymbol{K}\cdot(\boldsymbol{a}_{\perp}+\hat{\mathbf{z}})=0$
implies $-\Phi_{\text{Q}}=2d\Delta K_{z}$, the optical path due to
penetration of the FA states into the bulk. As a result, $\Phi_{\text{S}}-\Phi_{\text{Q}}=\Phi_{\textrm{FA}}$,
the total Berry phase from the FA states. We now calculate $\Phi_{\textrm{FA}}$
directly instead of splitting it as $\Phi_{\textrm{FA}}=\Phi_{\text{S}}+2\Delta\boldsymbol{K}\cdot\boldsymbol{d}$
and show that $\Phi_{\textrm{FA}}=\pi$ if $\mathcal{P}^{2}=-1$.
Explicitly, 
\begin{align}
\Phi_{FA} & =\intop_{\boldsymbol{K}_{\perp}^{1}}^{\boldsymbol{K}_{\perp}^{2}}i\left(\left\langle t(\boldsymbol{k})|\partial_{k_{FA}}t(\boldsymbol{k})\right\rangle -\left\langle b(\boldsymbol{k})|\partial_{k_{FA}}b(\boldsymbol{k})\right\rangle \right)dk_{FA}\nonumber \\
 & =\eta(\boldsymbol{K}_{\perp}^{2})-\eta(\boldsymbol{K}_{\perp}^{1})
\end{align}
using Eq.~(\ref{eq:P-top-bottom-1}), where $k_{\textrm{FA}}$ is
the momentum along the FA. A Berry phase generically is gauge-invariant
only for closed paths whereas the FAs are open contours. Thus, $\Phi_{\text{top}}=\intop_{\boldsymbol{K}_{\perp}^{1}}^{\boldsymbol{K}_{\perp}^{2}}i\left\langle t(\boldsymbol{k})|\partial_{k_{\textrm{FA}}}t(\boldsymbol{k})\right\rangle dk_{\textrm{FA}}$
and $\Phi_{\text{bottom}}=\intop_{\boldsymbol{K}_{\perp}^{2}}^{\boldsymbol{K}_{\perp}^{1}}i\left\langle b(\boldsymbol{k})|\partial_{k_{\textrm{FA}}}b(\boldsymbol{k})\right\rangle dk_{\textrm{FA}}$
are not gauge-invariant, which makes $\Phi_{\textrm{FA}}$ naïvely
ambiguous. To resolve this paradox, we note that each point on each
FA can be understood as an edge state of a Chern insulator. While
a single edge of a Chern insulator violates gauge invariance and exhibits
a 1D chiral anomaly, opposite edges together respect gauge invariance,
so $\Phi_{FA}$ is indeed gauge invariant.

To determine $\eta(\boldsymbol{K}_{\perp}^{2})-\eta(\boldsymbol{K}_{\perp}^{1})$,
we consider the action of $\mathcal{P}$ on the bulk states. Since
Weyl nodes at $\boldsymbol{K}^{1}$ and $\boldsymbol{K}^{2}$ have
opposite chiralities, an electron Fermi surface around $\boldsymbol{K}^{1}$,
carries the same Chern number as a hole Fermi surface that encloses
$\boldsymbol{K}^{2}$. As a result, a smooth set of unitary transformations
exists that deform Bloch states on the former into Bloch states on
the latter. Shrinking these Fermi surfaces to vanishing volume around
the Weyl nodes then reduces the unitary transformations to a pure
phase, $e^{i\alpha}$, which implies that an upward dispersing chiral
mode at $\boldsymbol{K}^{1}$ has the same Bloch ket as a downward
dispersing chiral mode at $\boldsymbol{K}^{2}$, and vice versa. Now,
the FA states smoothly merge with the bulk states at the Weyl nodes,
so the upward chiral modes at both $\boldsymbol{K}^{1}$ and $\boldsymbol{K}^{2}$
are simply the end-points of the bottom FA, $|b(\boldsymbol{K}^{1})\rangle$
and $|b(\boldsymbol{K}^{2})\rangle$, respectively, while the downward
modes are $|t(\boldsymbol{K}^{1})\rangle$ and $|t(\boldsymbol{K}^{2})\rangle$,
or vice-versa. Therefore, $|t(\boldsymbol{K}^{1})\rangle=e^{i\alpha}|b(\boldsymbol{K}^{2})\rangle$
and $|t(\boldsymbol{K}^{2})\rangle=e^{i\alpha}|b(\boldsymbol{K}^{1})\rangle$,
which yields 
\begin{align}
e^{i\eta(\boldsymbol{K}_{\perp}^{2})} & =\left\langle b(\boldsymbol{K}_{\perp}^{2})\left|\mathcal{P}\right|t(\boldsymbol{K}_{\perp}^{2})\right\rangle =\left\langle t(\boldsymbol{K}_{\perp}^{1})\left|\mathcal{P}\right|b(\boldsymbol{K}_{\perp}^{1})\right\rangle =-e^{i\eta(\boldsymbol{K}_{\perp}^{1})}
\end{align}
or $\Phi_{\textrm{FA}}=\eta(\boldsymbol{K}_{\perp}^{2})-\eta(\boldsymbol{K}_{\perp}^{1})=\pi$,
using Eq.~(\ref{eq:P-top-bottom-1}) with $\eta^{\prime}(\boldsymbol{k})=\eta(\boldsymbol{k})+\pi$,
which follows from $\mathcal{P}^{2}=-1$.

\section{Verification on a lattice model\label{sec:Lattice-verification}}

\begin{table*}
\begin{tabular}{p{0.2\textwidth}|p{0.35\textwidth}|p{0.35\textwidth}}
\hline
%{|>{\centering}m{0.1\textwidth}|>{\centering}m{0.3\textwidth}|>{\centering}m{0.3\textwidth}|}
Segment & Behavior in $\boldsymbol{\tau}\otimes\boldsymbol{\sigma}$ & Behavior in $\boldsymbol{S}$\tabularnewline
\hline 
\hline 
Top surface FA & Rotation of $\left\langle \boldsymbol{\sigma}\right\rangle $ about
$\sigma_{z}$ by $\phi$, with fixed $\left\langle \boldsymbol{\tau}\right\rangle =(0,0,1)$. & Rotation of $\left\langle \boldsymbol{S}\right\rangle $ about $S_{z}$
by $\phi$ with fixed $\left\langle S_{z}\right\rangle =1/2$.\tabularnewline
\hline 
Downward bulk travel & $\pi$ rotation of $\left\langle \boldsymbol{\tau}\right\rangle $
from $(0,0,1)$ to $(0,0,-1)$ with fixed $\left\langle \boldsymbol{\sigma}\right\rangle $. & Rotation of $\left\langle \boldsymbol{S}\right\rangle $ from $\left\langle S_{z}\right\rangle =+1/2$
to $\left\langle S_{z}\right\rangle =-1/2$ with fixed $\left\langle S_{x}\right\rangle ,\left\langle S_{y}\right\rangle $.\tabularnewline
\hline 
Bottom surface FA & Rotation of $\left\langle \boldsymbol{\sigma}\right\rangle $ about
$\sigma_{z}$ by $-\phi$, with fixed $\left\langle \boldsymbol{\tau}\right\rangle =(0,0,-1)$. & Rotation of $\left\langle \boldsymbol{S}\right\rangle $ about $S_{z}$
by $\phi$ with fixed $\left\langle S_{z}\right\rangle =-1/2$.\tabularnewline
\hline 
Upward bulk travel & $\pi$ rotation of $\left\langle \boldsymbol{\tau}\right\rangle $
from $(0,0,-1)$ to $(0,0,1)$ with fixed $\left\langle \boldsymbol{\sigma}\right\rangle $. & Rotation of $\left\langle \boldsymbol{S}\right\rangle $ from $\left\langle S_{z}\right\rangle =-1/2$
to $\left\langle S_{z}\right\rangle =+1/2$ with fixed $\left\langle S_{x}\right\rangle ,\left\langle S_{y}\right\rangle $.\tabularnewline
\hline 
\end{tabular}

\caption{Description of the four segments of a semiclassical orbit in bilayer-spin
space, $\boldsymbol{\tau}\otimes\boldsymbol{\sigma}$, and in terms
of total spin, $\boldsymbol{S}=(\boldsymbol{\tau}+\boldsymbol{\sigma})/2$.
The angle $\phi$ is given in Eq.~(\ref{eq:spin-rotation}).}
\end{table*}

\subsection{Bloch Hamiltonian and Fermi arcs \label{subsec:Lattice-model}}

Eq.~(\ref{eqn:TWSM_full}) below defines the bulk Bloch Hamiltonian
and spectrum of the lattice model, which contains two layers per unit
cell along $z$, taken to be the surface normal:

\begin{eqnarray}
H_{0}\left(\boldsymbol{k},k_{z}\right) & = & \boldsymbol{\tau}\cdot\boldsymbol{d}\left(\boldsymbol{k},k_{z}\right)-\mu\label{eqn:TWSM_full}\\
\varepsilon_{\pm}^{2}\left(\boldsymbol{k},k_{z}\right) & = & \left(\sqrt{v_{x}^{2}\sin^{2}k_{x}+v_{y}^{2}\sin^{2}k_{y}}\pm\ell\right)^{2}+d_{\perp}^{2}(\boldsymbol{k},k_{z})\nonumber 
\end{eqnarray}
where $\boldsymbol{k}=(k_{x},k_{y})$, $\tau_{z}=\pm1$ for the two
layers of the bilayer, $d_{x}\left(\boldsymbol{k},k_{z}\right)=m_{0}+\sum_{i=x,y,z}\beta_{i}\cos k_{i}$,
$d_{y}\left(\boldsymbol{k},k_{z}\right)=\sum_{i=x,y,z}u_{i}\sin k_{i}$,
$d_{\perp}^{2}(\boldsymbol{k},k_{z})=|d_{x}(\boldsymbol{k},k_{z})|^{2}+|d_{y}(\boldsymbol{k},k_{z})|^{2}$
and $d_{z}\left(\boldsymbol{k},k_{z}\right)\equiv d_{z}(\boldsymbol{k})=\sigma_{x}v_{x}\sin k_{x}+\sigma_{y}v_{y}\sin k_{y}-\ell$
denotes purely in-plane hopping that captures contains spin-orbit
coupling through spin Pauli matrices $\sigma_{x,y}$ and $\mathcal{I}$
symmetry breaking through the term $\propto\ell$. It preserves $\mathcal{T}$
symmetry ($\mathcal{T}=\sigma_{y}\mathbb{K}$) but breaks all spatial
symmetries for general $(u_{x},u_{y},u_{z})$. It preserves a chiral
symmetry, $\tilde{I}=\tau_{y}\sigma_{z}\otimes(\boldsymbol{r}\to-\boldsymbol{r})$,
which is better understood as spatial inversion about a point between
the layers of the bilayer, $\tau_{x}\sigma_{z}\otimes(\boldsymbol{r}\to-\boldsymbol{r})$,
followed by a local unitary transformation $\psi\to e^{i\tau_{z}\sigma_{z}\frac{\pi}{2}}\psi$.
The resulting particle-hole symmetry, $\mathcal{P}=\mathcal{T}\tilde{I}$,
causes FAs on opposite surfaces to coincide.

The prescription yields bulk WNs when $\varepsilon_{-\text{sgn}(\ell)}(\boldsymbol{k})=0$,
while surface FAs occur between projections of the WNs along curves
where $d_{z}(\boldsymbol{k})$ has zero eigenvalues. For $u_{x}=u_{y}=0$,
all the nodes lie at either $k_{z}=0$ or $\pi$, while non-zero $u_{x},u_{y}$
place the WNs to distinct $k_{z}$. In our calculations, we choose
band parameters $\left\{ v_{x},v_{y},\ell\right\} =\left\{ 3.53,2.48,3.00\right\} $,
$\left\{ m_{0},\beta_{x},\beta_{y},\beta_{z}\right\} =\left\{ 1.000,-0.939,0.371,0.652\right\} $,
$\left\{ u_{x},u_{y},u_{z}\right\} =\left\{ u\cos\pi/5,u\sin\pi/5,-1\right\} $,
and tune $u$ to create different WN and FA configurations.

\subsection{$\Phi_{\text{S}}$ as a Berry phase on a Bloch sphere}

\label{subsec:PhiS}

\begin{figure}
\includegraphics[width=0.5\columnwidth]{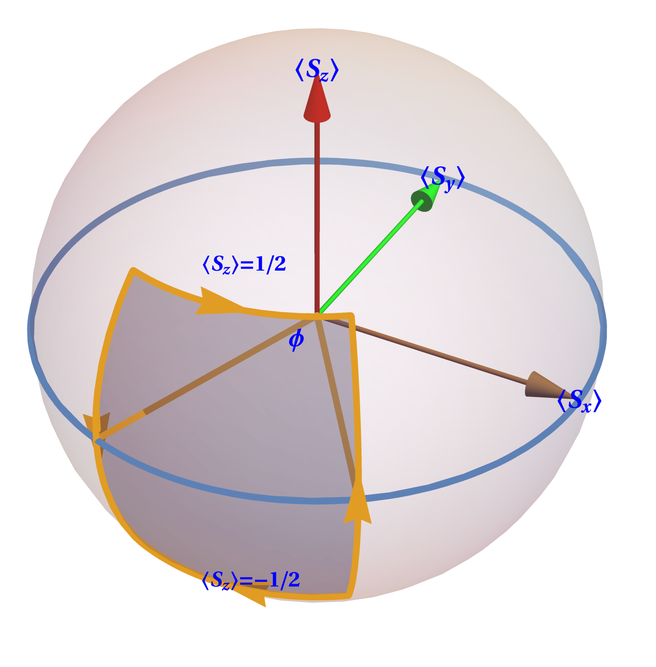}\caption{Semiclassical orbit (yellow curve) on the $S=1$ Bloch sphere that
results from the triplet combination of the spin ($\sigma$) and bilayer
pseudospin ($\tau$) degrees of freedom. Horizontal (vertical) arms
of the orbit capture motion along the FAs (through the bulk) and yield
a Berry phase equal to the solid angle enclosed by the orbit. \label{fig:Bloch sphere}}
\end{figure}

The lattice model defined by Eq.~(\ref{eqn:TWSM_full}) admits an
elegant analytical determination of $\Phi_{\text{S}}$, which facilitates
comparison with the numerics. We calculate this first.

As stated above, FAs occur along curves where $d_{z}(\boldsymbol{k})$
has zero eigenvalues. Thus, $(k_{x},k_{y})$ satisfy $v_{x}^{2}\sin^{2}k_{x}+v_{y}^{2}\sin^{2}k_{y}=\ell^{2}$
along such a curve while the spin part of the wavefunction is an eigenstate
of $\sigma_{x}v_{x}\sin k_{x}+\sigma_{y}v_{y}\sin k_{y}$ with eigenvalue
$\ell$. As a result, the surface projection of a FA state at $\boldsymbol{k}$,
$|\psi_{\boldsymbol{k},\gamma}^{\textrm{FA}}\rangle$, satisfies $\left\langle \psi_{\boldsymbol{k},\gamma}^{\textrm{FA}}\left|\sigma_{i}\right|\psi_{\boldsymbol{k},\gamma}^{\textrm{FA}}\right\rangle =v_{i}\sin k_{i}/\ell$,
where $\gamma=\pm1$ denote FAs on the top (bottom) surface, while
the spin rotates by an angle 
\begin{equation}
\phi=\arg\left(\frac{v_{x}\sin K_{x}^{2}+iv_{y}\sin K_{y}^{2}}{v_{x}\sin K_{x}^{1}+iv_{y}\sin K_{y}^{1}}\right)\label{eq:spin-rotation}
\end{equation}
along a FA that connects surface projections of WNs at $\boldsymbol{K}^{1}$
and $\boldsymbol{K}^{2}$. Moreover, $|\psi_{\boldsymbol{k},\gamma}^{\textrm{FA}}\rangle$
satisfy 
\begin{equation}
\left[\tau_{x}d_{x}(\boldsymbol{k},-i\partial_{z})+\tau_{y}d_{y}(\boldsymbol{k},-i\partial_{z})\right]|\psi_{\boldsymbol{k},\gamma}^{\textrm{FA}}\rangle=0
\end{equation}
which immediately implies that $|\psi_{\boldsymbol{k},\gamma}^{\textrm{FA}}\rangle$
are Jackiw-Rebbi zero modes in bilayer space, spanned by $\tau_{i}$.
In particular, they are eigenstates of $\tau_{z}$ with opposite eigenvalues
on the top and bottom surface. To determine $\Phi_{\mathrm{S}}$,
we view the $\sigma$ and $\tau$ degrees of freedom as two spin-1/2
particles, and consider the effect of the above rotations on the total
spin, $\boldsymbol{S}=(\boldsymbol{\tau}+\boldsymbol{\sigma})/2$.
Table I describes the four segments of a semiclassical orbit involving
WNs at $\boldsymbol{K}_{1}$ and $\boldsymbol{K}_{2}$ in terms of
both $\boldsymbol{\tau}\otimes\boldsymbol{\sigma}$ and $\boldsymbol{S}$.

On the Bloch sphere for total spin, these segments define a patch
with area $\phi\times\left[\frac{1}{2}-\left(-\frac{1}{2}\right)\right]=\phi$,
as shown in Fig.~\ref{fig:Bloch sphere}. This patch induces a Berry
phase $S\phi$, which vanishes in the singlet sector ($S=0$) and
equals $\phi$ in the triplet sector ($S=1$). Thus, the total Berry
phase acquired through the above rotations is 
\begin{equation}
\Phi_{\text{S}}=\phi
\end{equation}
where $\phi$ is given in Eq.~(\ref{eq:spin-rotation}).

\subsection{Fitting of zero modes}

To resolve the Berry phase dependence of the zero point energy, we
note that Eq.~(1) of the main paper predicts a pair of zero modes,
as seen for the $y$-leaning vortex, whenever $\Phi_{\text{tot}}=\Phi_{\text{B}}+\Phi_{\text{S}}-\Phi_{\text{Q}}$
equals an odd multiple of $\pi$. In this geometry, $\Phi_{\text{B}}=(\Delta K_{y}a_{y}+\Delta K_{z})L_{z}$,
$\Phi_{\text{S}}$ is determined by the the band structure as described
in Sec.~\ref{subsec:PhiS} and is $a_{y}$- and $L_{z}$-independent,
and $\Phi_{\text{Q}}\propto a_{y}$ and is $L_{z}$-independent. An
immediate consequence is that $\Phi_{\text{tot}}$ becomes independent
of $L_{z}$ at the ``magic angle'', $\theta=-\tan^{-1}(\Delta K_{z}/\Delta K_{y})$.
Interestingly, we see in Fig.~\ref{fig:fits}(a) that zero modes
exist precisely at the magic tilt, $a_{y}\approx0.424$, implying
that $\Phi_{\text{S}}-\Phi_{\text{Q}}=\pi\mod2\pi$ at this tilt.
This is consistent with our prediction in Sec.~\ref{subsec:MM-SUSY-magic}
that magic angle vortices are critical if FAs on opposite surface
coincide in the normal state.

\begin{figure*}
\begin{centering}
\subfloat[]{\includegraphics[width=0.45\textwidth]{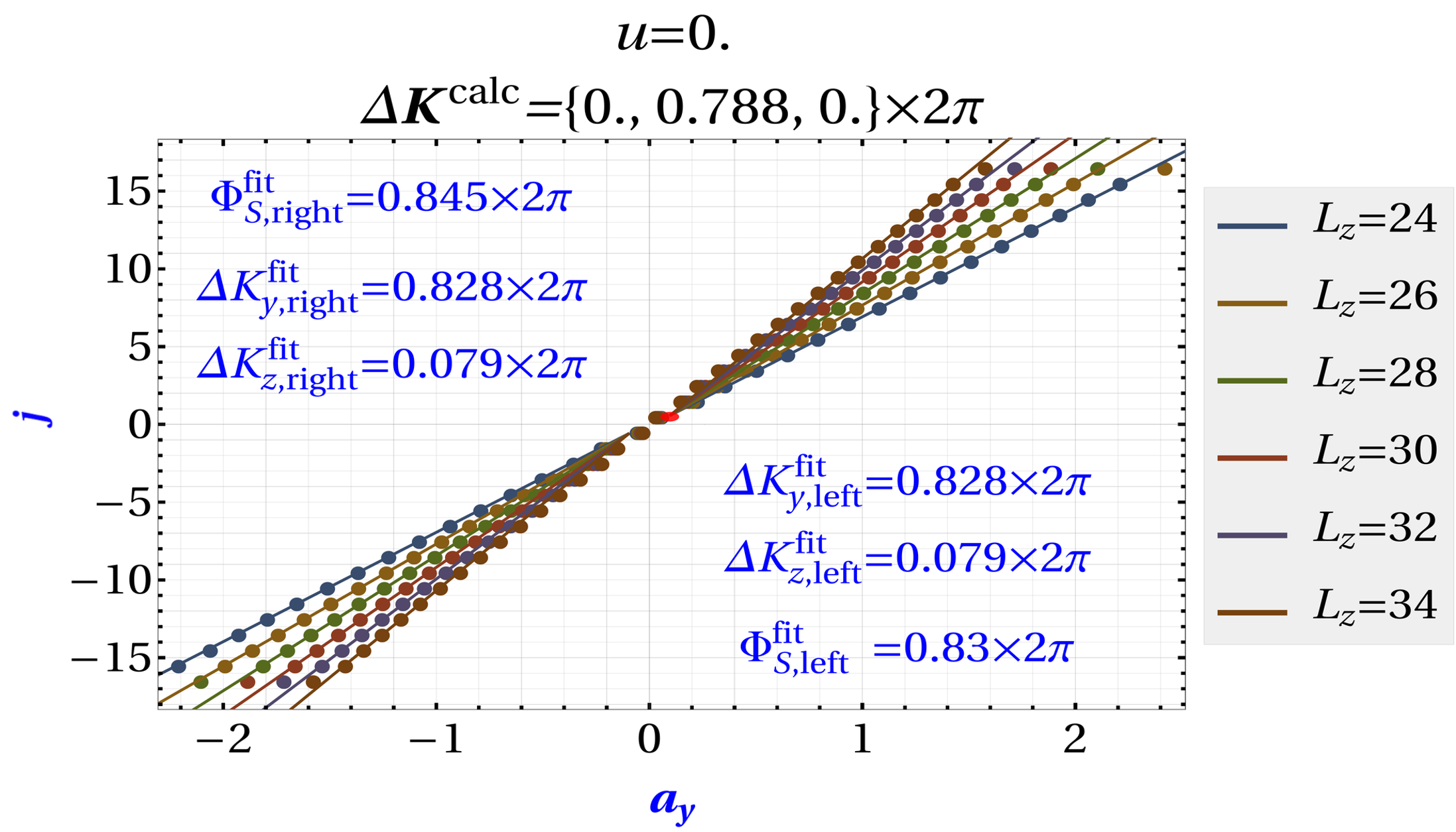}

}\subfloat[]{\includegraphics[width=0.45\textwidth]{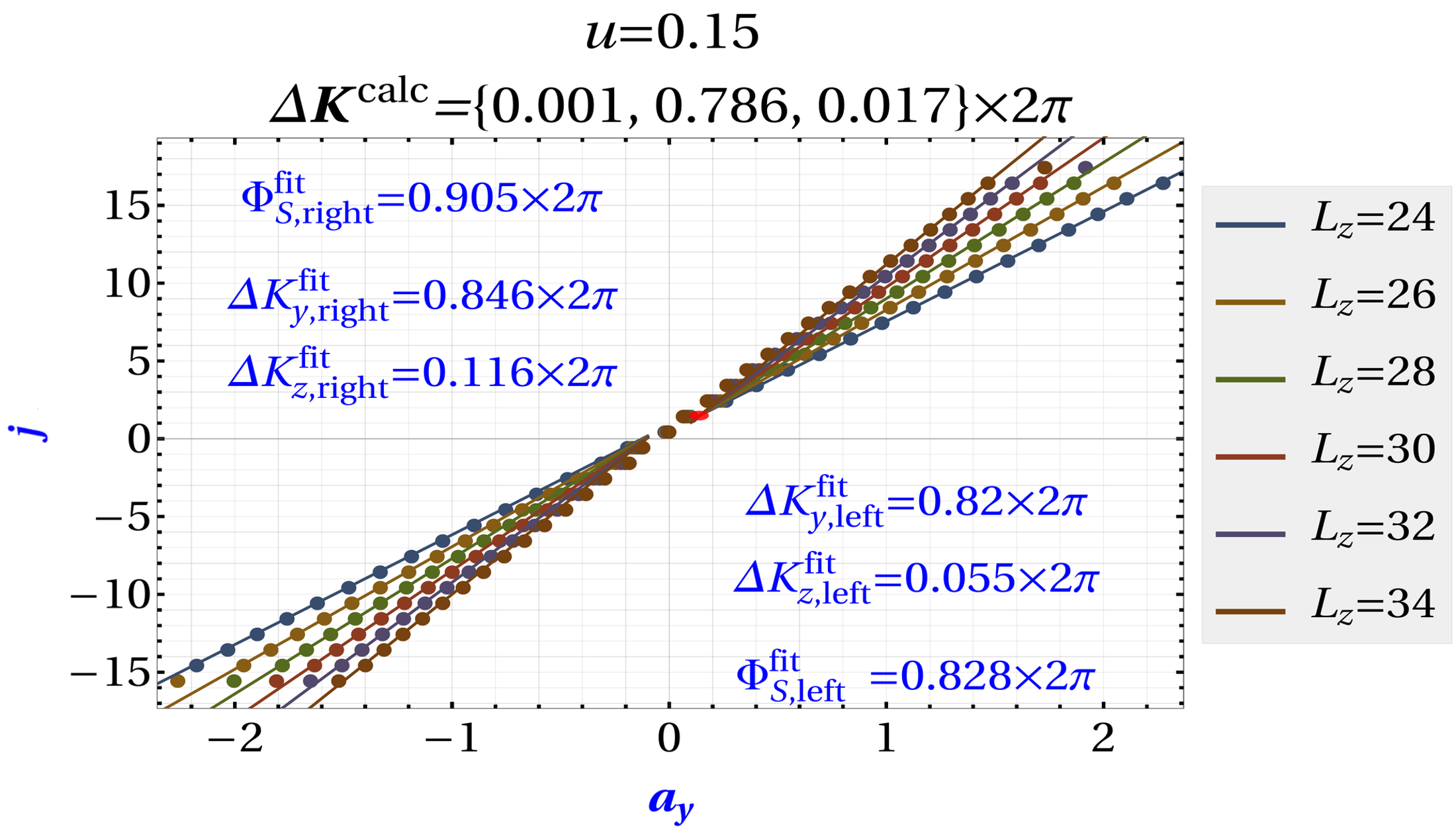}

}
\par\end{centering}
\begin{centering}
\subfloat[]{\includegraphics[width=0.45\textwidth]{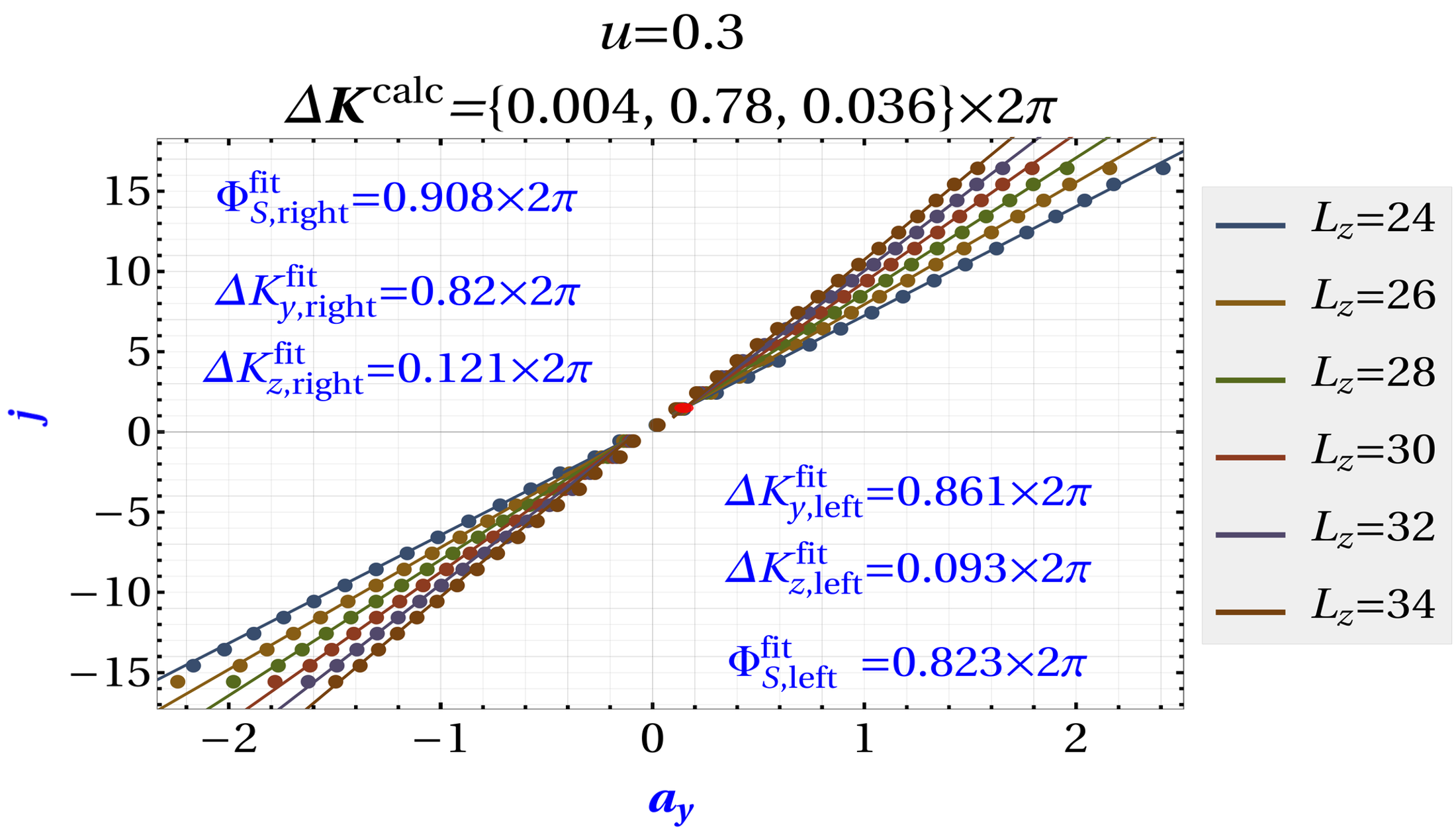}

}\subfloat[]{\includegraphics[width=0.45\textwidth]{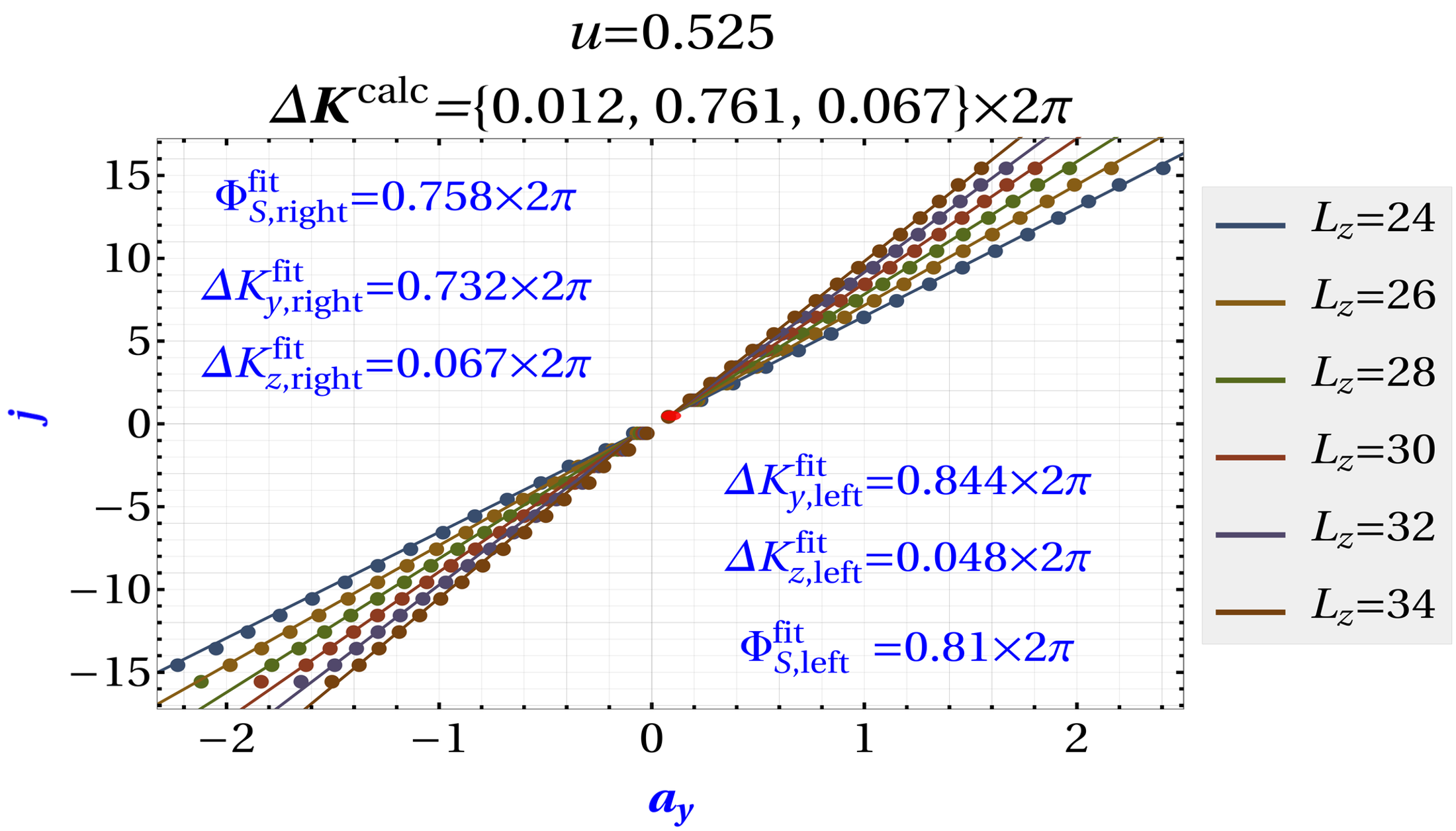}

}
\par\end{centering}
\begin{centering}
\subfloat[]{\includegraphics[width=0.45\textwidth]{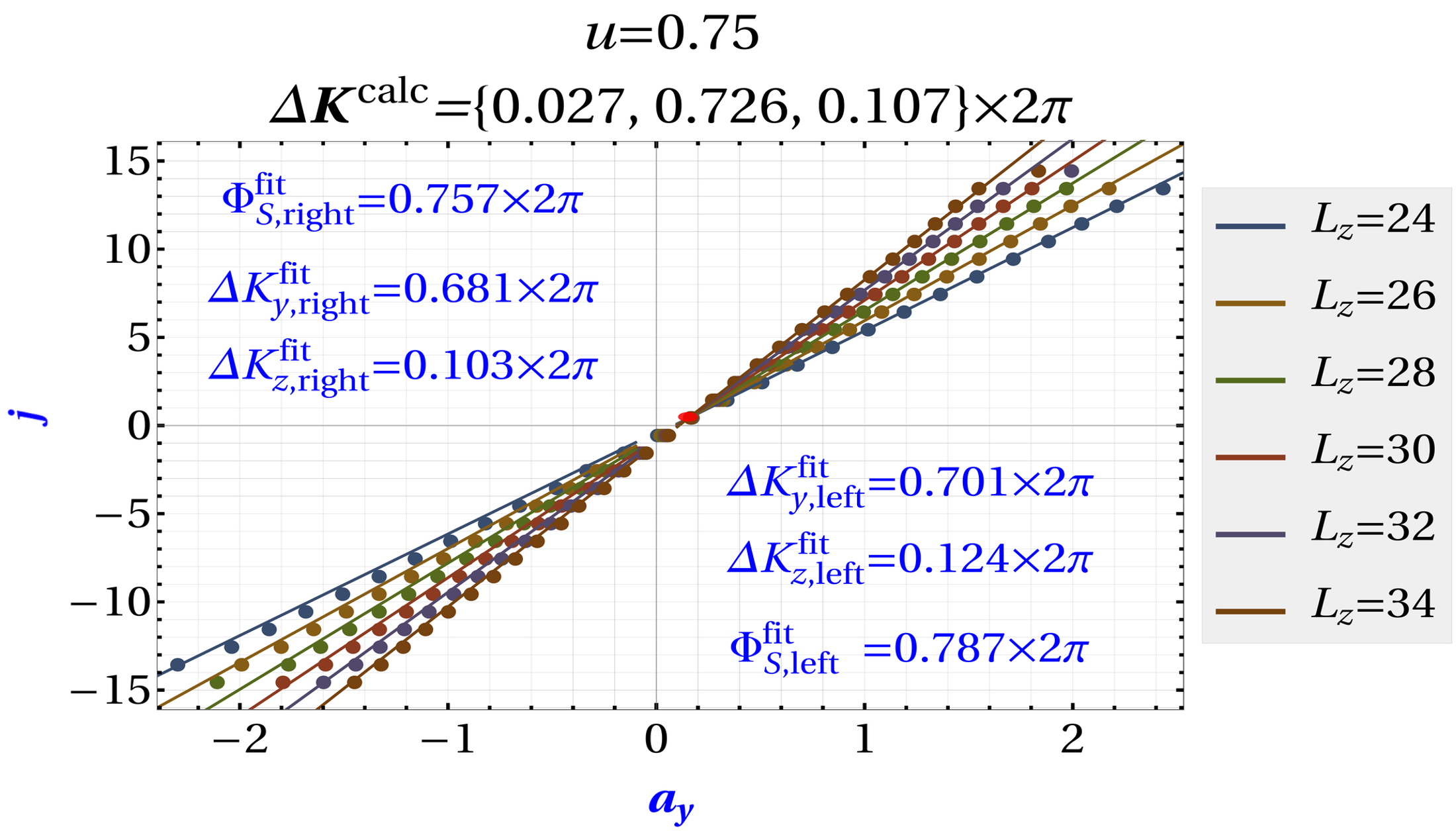}

}\subfloat[]{\includegraphics[width=0.45\textwidth]{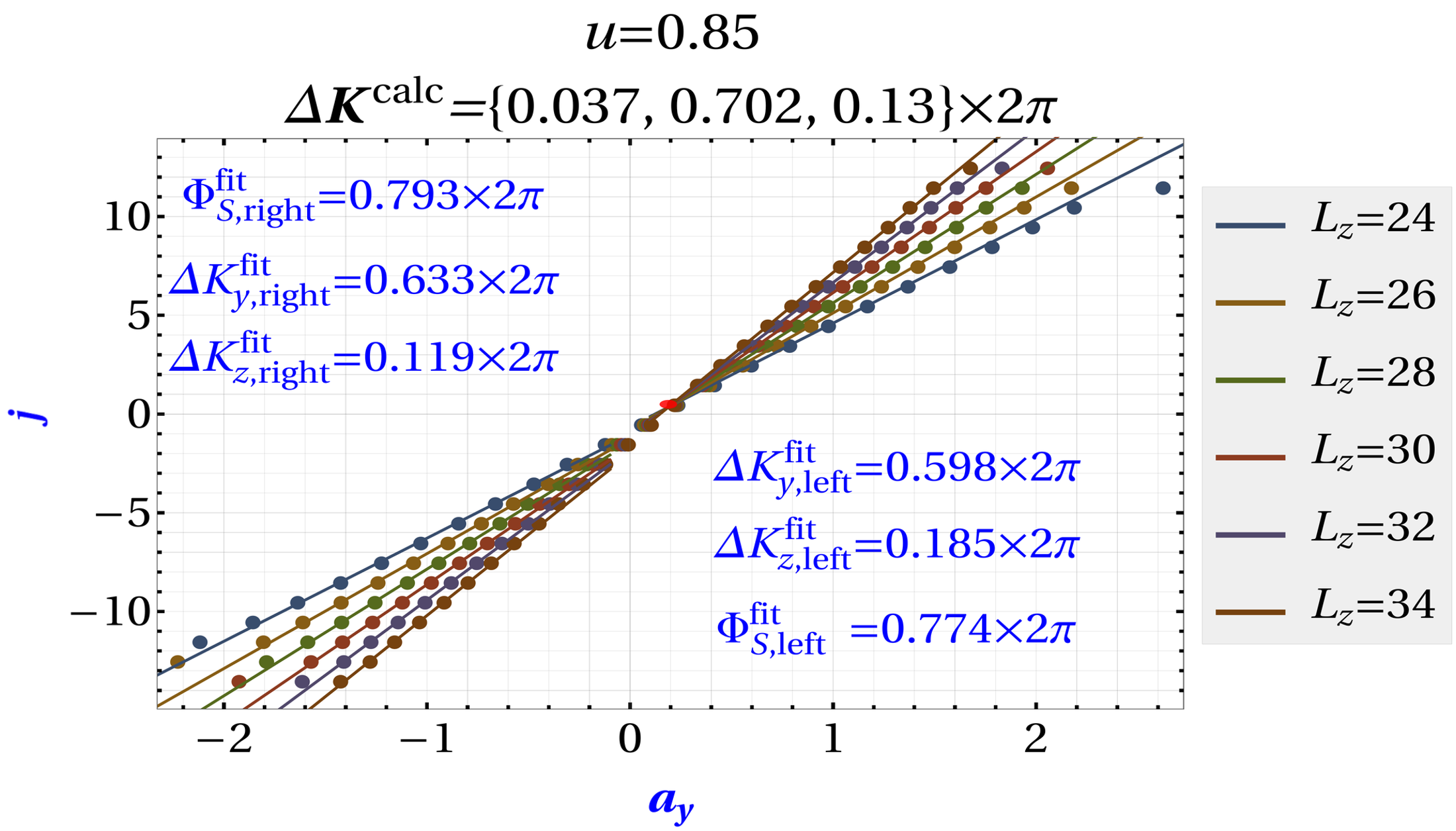}

}
\par\end{centering}
\begin{centering}
\subfloat[]{\includegraphics[width=0.45\textwidth]{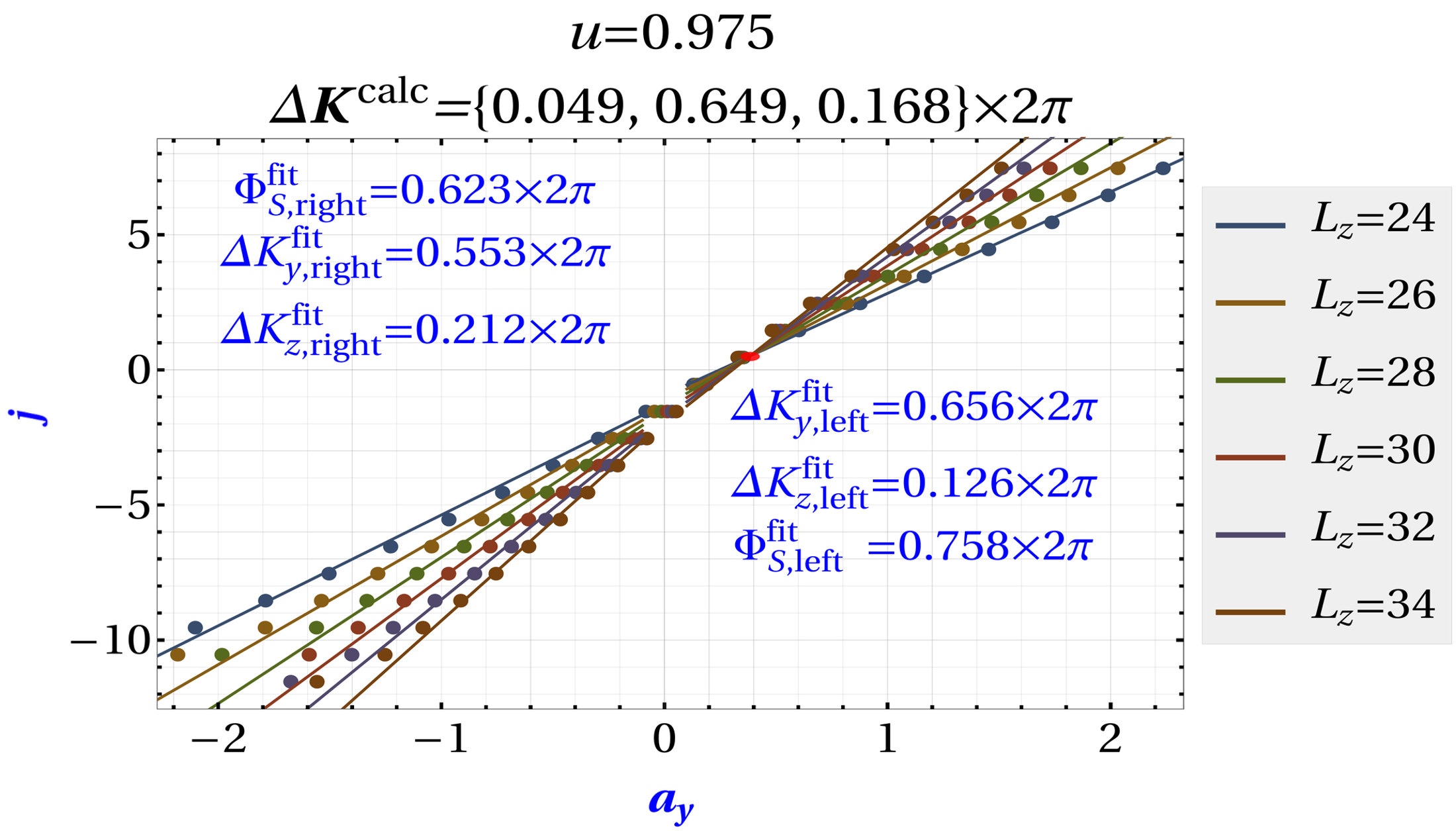}

}\subfloat[]{\includegraphics[width=0.45\textwidth]{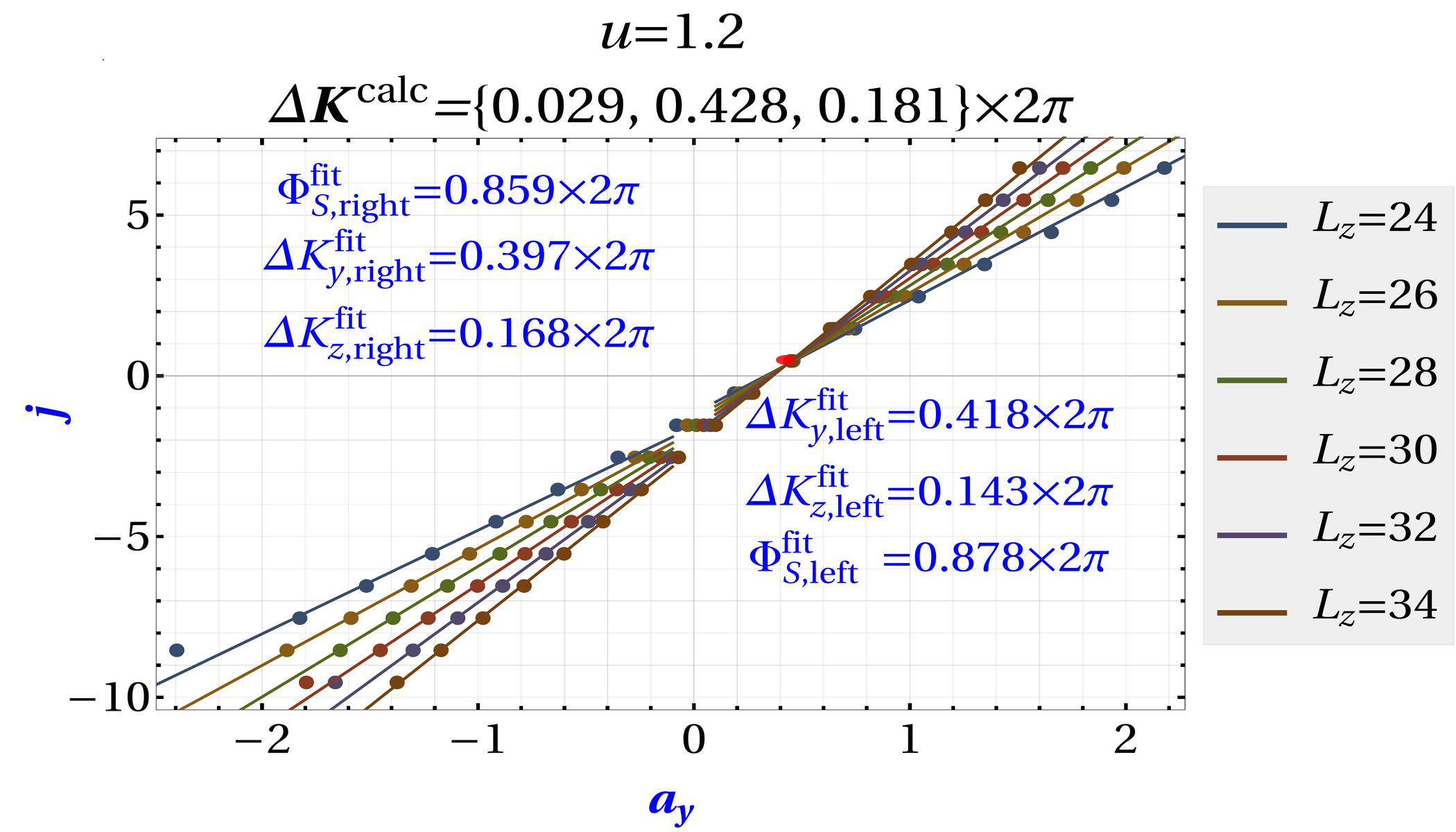}

}
\par\end{centering}
\caption{(a-h) Zero modes indices ($j$) vs $a_{y}$ for different values of
the band parameter $u$. The points fit well to straight lines for
each $u$ and each $L_{z}$, but separately for $a_{y}>0$ (\textquotedblleft right\textquotedblright )
and $a_{y}<0$ (\textquotedblleft left\textquotedblright ). At each
$u$, all the lines intersect at a certain value of $a_{y}$, which
defines the magic tilt. The symmetries of the model guarantee zero
modes precisely at this tilt.\label{fig:fits}}
\end{figure*}

\begin{figure*}
\begin{centering}
\subfloat[]{\includegraphics[width=0.5\textwidth]{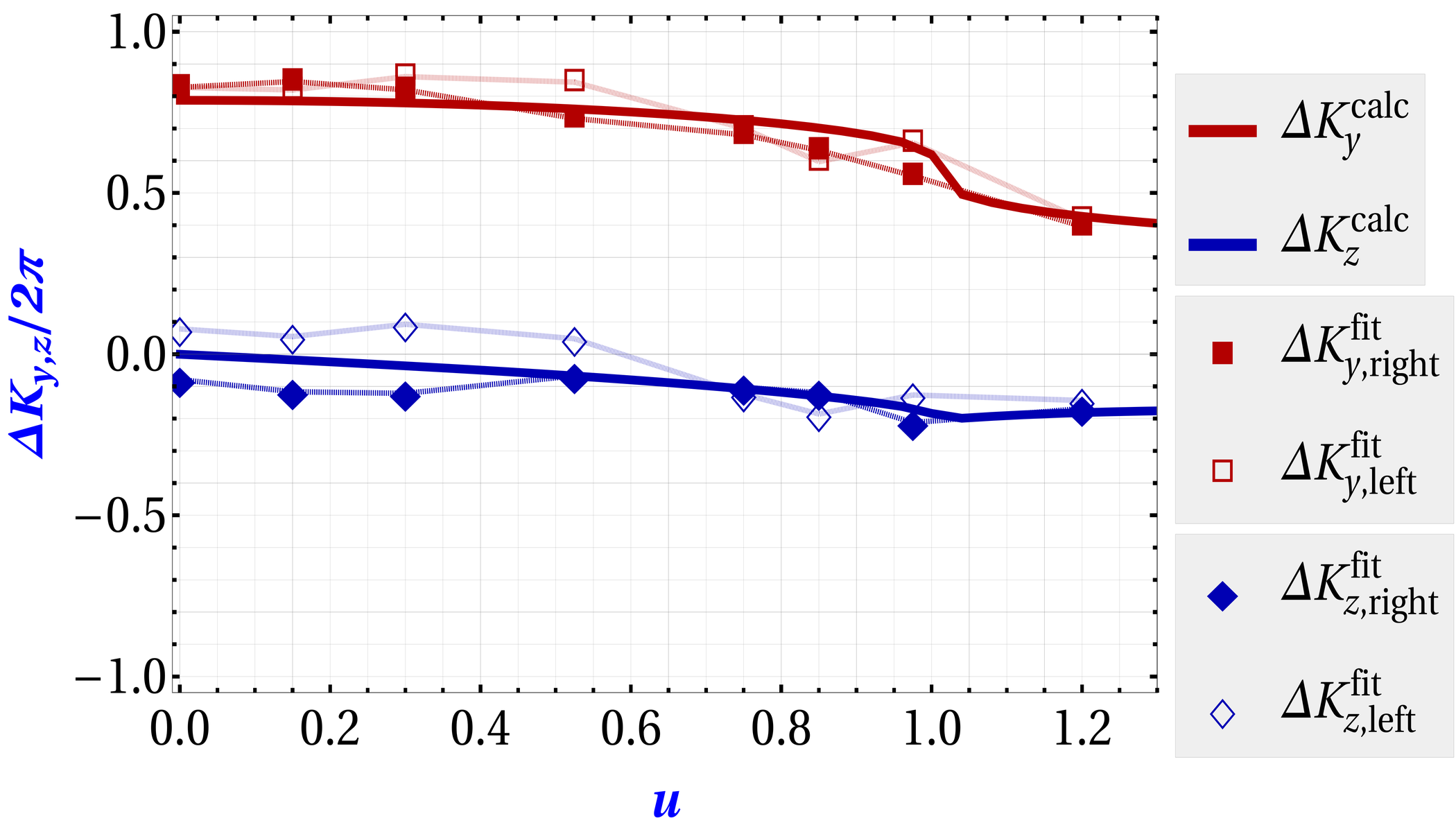}

}\subfloat[]{\includegraphics[width=0.5\textwidth]{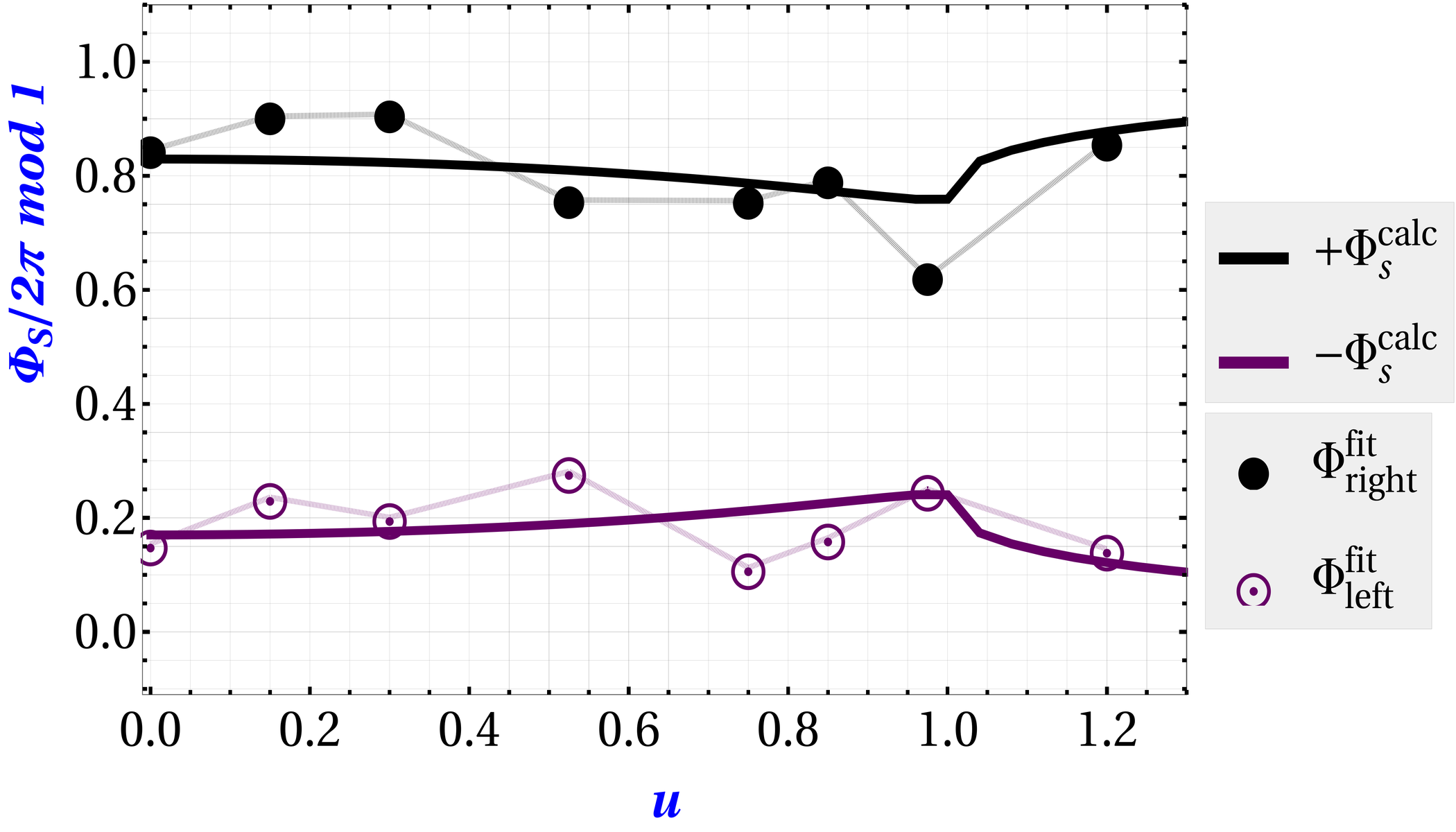}

}
\par\end{centering}
\caption{(a) The calculated and fitted values of $\Delta K_{y,z}$ are similar
and in good agreement and for both $a_{y}>0$ and $a_{y}<0$ for a
wide range of band parameters, parameterized by $u$. (b) The calculated
and fitted values of $\Phi_{\text{S}}$ are in good agreement, and
are equal and opposite ($\mod2\pi$) for $a_{y}>0$ and $a_{y}<0$.\label{fig:fits-1}}
\end{figure*}

\begin{figure}
\includegraphics[width=0.6\columnwidth]{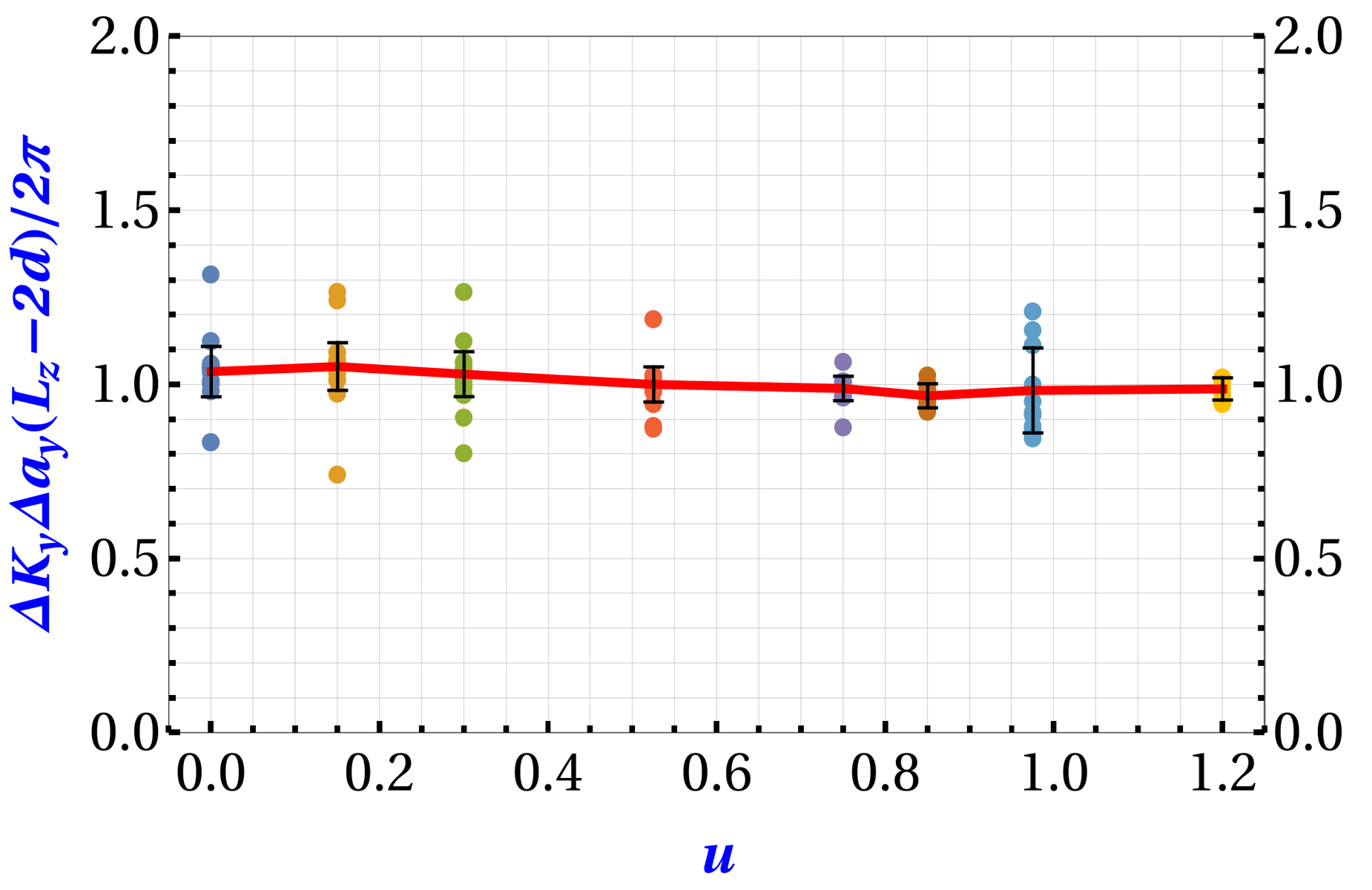}

\caption{Verification of Eq.~(4) of the main paper at $a_{x}=0$ for various
$u$ when $L_{z}=34$. Each dot at a given $u$ denotes a suitably
scaled value of $\Delta a_{y}$ for a pair of consecutive peaks. Clearly,
the mean $\frac{\left\langle \Delta K_{y}\Delta a_{y}\right\rangle }{2\pi/(L_{z}-2d)}\approx1$
in accordance with Eq.~(4) for each $u$ with a small standard deviation,
denoted by error bars.\label{fig:fits-2}}
\end{figure}

We proceed to fit $\Phi_{\text{tot}}$ to a suitable function of $a_{y}$
and $L_{z}$ and extract the values of $\Delta K_{y}$, $\Delta K_{z}$
and $\Phi_{\text{S}}$. Specifically, for fixed band parameters in
the normal state, we set $a_{x}=0$ and vary $a_{y}$ between $\pm L_{y}/L_{z}$;
for larger $a_{y}$, the vortex enters and exits from the side surfaces.
Zero modes exist at regular intervals of $a_{y}$, which we expect
to correspond to $\Phi_{\text{B}}+\Phi_{\text{S}}-\Phi_{\text{Q}}$
sweeping past an odd multiple of $\pi$. Defining $j=\Phi_{\text{tot}}/2\pi$,
we assign consecutive half-integer $j$ values to the zero modes for
fixed $L_{z}$. For each $L_{z}$, $\Phi_{\text{tot}}\left(L_{z}\right)$
fits excellently to separate straight lines for $a_{y}>0$ (right
tilt) and $a_{y}<0$ (left tilt):

\begin{equation}
\Phi_{\text{fit}}\left(L_{z},a_{y}\right)=m\left(L_{z}\right)a_{y}+c\left(L_{z}\right)
\end{equation}
Fitting must be performed separately for $a_{y}>0$ and $a_{y}<0$
because the semiclassical orbits for the two cases encircle the vortex
in opposite directions and acquire equal and opposite $\Phi_{S}$,
but yield the same values for the other parameters. Moreover, zero
modes near $a_{y}=0$ must be ignored because they involve interference
between clockwise and counterclockwise orbits around the vortex, which
causes deviations from the semiclassical limit. We also ignore zero
modes for large $|a_{y}|$, when the vortex ends are near the edge
of the lattice. The slope and intercept, $m(L_{z})$ and $c(L_{z})$,
are each found to be almost perfect straight lines functions of $L_{z}$.
The upshot is that $\Phi_{\text{fit}}$ is of the form 
\begin{equation}
\Phi_{\text{fit}}(L_{z},a_{y})=A+Ba_{y}+CL_{z}+Da_{y}L_{z}\label{eq:Phi-fit}
\end{equation}
while we expect 
\begin{equation}
\Phi_{\text{tot}}(L_{z},a_{y})=\Phi_{S}-2\Delta K_{y}da_{y}+\Delta K_{z}L_{z}+\Delta K_{y}a_{y}L_{z}\label{eq:Phi-tot}
\end{equation}
Comparing (\ref{eq:Phi-fit}) and (\ref{eq:Phi-tot}), we extract
the values of $\Delta K_{y}$, $\Delta K_{z}$, $\Phi_{\text{S}}$
and $d$. As is evident from Figs.~\ref{fig:fits-1} and \ref{fig:fits-2},
the first three parameters match remarkably well with values calculated
directly in the normal state with small errors indicating good fits,
while $d$ gives $\Phi_{\text{S}}+2d\Delta K_{z}\approx\pi$ at the
magic angle as expected.
\end{widetext}

\bibliographystyle{unsrt}
\bibliography{library}

\end{document}